\documentclass{article}
\usepackage{authblk}
\usepackage{hyperref}
\usepackage{float}
\usepackage{xcolor} 
\usepackage{stfloats}
\usepackage{placeins}  
\usepackage{ragged2e}
\usepackage{amsmath}
\usepackage{graphicx}
\usepackage{caption}
\usepackage{epstopdf,epsfig}
\usepackage{newtxtext}
\usepackage{newtxmath}
\usepackage{natbib}
\usepackage{hyperref}
\usepackage[utf8]{inputenc}
\usepackage{cleveref}
\usepackage[a4paper, margin=3cm]{geometry}
\date{}

\title{Data-driven transient growth analysis}

\author[1]{Yin Wang}
\author[2,3]{Xuerui Mao\thanks{Corresponding author: maoxuerui@sina.com}}

\affil[1]{School of Mechatronical Engineering, Beijing Institute of Technology, Beijing 100081, P.R. China}
\affil[2]{Beijing Institute of Technology (Zhuhai), Zhuhai, 519088, China}
\affil[3]{State Key Laboratory of Explosion Science and Safety Protection, Beijing, 100081, China}

\begin{document}
	\maketitle
	
	\begin{abstract}
		Transient growth analysis has been extensively studied in asymptotically stable flows to identify their short-term amplification of perturbations. Generally, in global transient growth analyses, matrix-free methods are adopted, requiring the construction of adjoint equations, either in the discrete or continuous form. This paper introduces a data-driven algorithm that circumvents the adjoint equations by extracting the optimal initial perturbation and its energy growth over a specified time horizon from transient snapshots of perturbations. This method is validated using data from the linearised complex Ginzburg-Landau equation, backward-facing step flow, and the Batchelor vortex. Unlike model-based methods, which require $S$ sets of integrations of the linearised governing equation and its adjoint for $S$ time horizons, the proposed approach collects the snapshots of $S$ time horizons in one integration of the linearised equation. Furthermore, this study provides a robust framework for utilising proper orthogonal decomposition (POD) modes to synthesise optimal modes. The developed capacity to conduct transient growth analyses without solving the adjoint equations is expected to significantly reduce the barriers to transient dynamics research.
		
	\end{abstract}
	
	\section{Introduction}
	\label{sec: Introduction}
	
	Perturbation analyses investigate the response of a dynamical system to infinitesimally small disturbances, providing insights into its short- and long-term behaviours. Classical linear stability theory examines the eigenvalue spectrum of the governing operator to characterise the asymptotic fate of perturbations, with eigenvalues determining whether disturbances decay or grow exponentially over infinite time horizons \citep{schmidpj2001stabilityandtransitionin}. 
	
	However, significant limitations of asymptotic stability have been identified in practical applications. Experimental and numerical investigations have demonstrated that eigenvalue analyses fail to adequately describe the dynamics of disturbances. These failures were initially attributed to the nonlinear dynamics but were then found to be due to the transient growth of perturbations over a short time interval, which can be substantial even in asymptotically stable flows \citep{farrell1988optimal, trefethen1993hydrodynamic}. Further studies demonstrated that transient growth is essential for understanding the sub-critical transition to turbulence in channel flow \citep{reddy1993energy}, bypass transition to turbulence in boundary layer flow \citep{andersson1999optimal}, etc. Consequently, transient growth analyses, focusing on the finite-time amplification of disturbances, have gained attention and offered a more comprehensive understanding of fluid physics. A range of physical mechanisms was then attributed to transient dynamics. For instance, the ``lift-up” effect describes the transient generation of streamwise velocity streaks via the transfer of momentum across the boundary layer by streamwise vortices \citep{landahl1980note}. Similarly, the Orr mechanism elucidates the process by which upstream-oriented perturbations are reoriented into downstream structures under the influence of base shear, thereby amplifying the disturbance energy \citep{farrell1988optimal}. The anti-lift-up mechanism, in contrast, explains the transient formation of toroidal structures of azimuthal vorticity induced by azimuthal velocity streaks at the periphery of forced vortices \citep{antkowiak2007vortex}.
	
	From the mathematical perspective, the pseudo-spectrum analysis proves that transient growth is closely tied to the non-normality of the linearised operator, which enables eigenmodes to interact and produce significant short-term growth \citep{trefethen1993hydrodynamic,tumin2001spatial,abdessemed2009linear}. Optimal initial perturbations can therefore be interpreted as the superposition of eigenmodes, as demonstrated in \citet{schmidpj2001stabilityandtransitionin}. These theoretical advancements were then extended to canonical flows such as the boundary layer flow \citep{obrist2003linear} and the Batchelor vortex \citep{mao2012transient}, which reveal that transient growth is often dominated by the continuous part of the spectrum. 
	
	Similar to asymptotic instability, the computation of transient growth can be categorised into two branches: direct and iterative methods \citep{reddy1993energy, barkley2008direct}. The former requires the explicitly discretized matrix form of the evolution operator of perturbations and is only numerically feasible for low-dimensional problems, such as those in local studies with base flows that are homogeneous in two directions. In the singular value decomposition (SVD) of the evolution operator, the largest singular value corresponds to the square root of the optimal energy growth, while the corresponding singular vectors provide the optimal initial perturbation and its outcome \citep{schmidpj2001stabilityandtransitionin}. In more general scenarios, particularly for global analyses involving base flows that are inhomogeneous in multiple directions, the matrix forms of evolution operators are practically unavailable. For instance, the discretisation of incompressible Navier--Stokes equations over a two-dimensional domain with moderate-resolution (e.g. the backward facing step flow presented in this work) typically leads to state vectors of dimension $\emph{O}\!\left(10^{4} \sim 10^{5}\right)$, while three-dimensional configurations can exceed $\emph{O}\!\left(10^{6}\right)$. As such, the full evolution operator would contain $\sim \emph{O}\!\left(10^{12}\right)$ entries, requiring several terabytes of memory just to store, even before any computations are performed. In such conditions, iterative methods, including eigenvalue and optimisation approaches, which are mathematically equivalent \citep{mao2013calculation}, are employed. The former involves constructing a Krylov sequence by iteratively evolving perturbations forwards using the linearised operator and backwards using its adjoint, followed by extracting the dominant eigenvalue and eigenvector of the joint operator \citep{barkley2008direct}. The latter determines the optimal initial perturbation by maximising a Lagrangian functional constrained by governing equations, again involving the iterative integration of the linearised operator and its adjoint \citep{schmid2007nonmodal}. 
	
	The analytical form of the adjoint equations can be derived using integration by parts \citep{barkley2008direct} and is thus called the continuous adjoint. Although widely used in flow stability studies, this form introduces additional boundary terms that complicate implementation, particularly for compressible flows \citep{raposo2019adjoint,de2012efficient} or in boundary perturbation analyses \citep{mao2015effects}. Conversely, the discrete adjoint approach derives the adjoint operator directly as the transconjugate of the discretised forward operator. This approach ensures machine-precision consistency between the forward and adjoint operators, rendering it particularly advantageous for gradient-based iterative methods, but it involves complex numerical implementation of boundary conditions for the adjoint. Considering the reliance on the adjoint for both the discrete and continuous approaches, we propose a data-driven methodology that captures transient dynamics using only flow field snapshots. By bypassing adjoint equations, this data-driven approach simplifies the analysis process and provides a practical alternative for transient growth studies.
	
	Data-driven analyses of perturbations have been extensively applied in fluid mechanics. One of the well-established techniques is the dynamic mode decomposition (DMD) \citep{schmid2010dynamic}, which approximates the eigenvalues and eigenfunctions of a dynamical system from a set of snapshots \citep{wei2009low, kutz2016dynamic}. Over the past decade, DMD has been used to analyse a wide variety of flows, and numerous variations of its original scheme have been developed. These improvements have enabled DMD to handle non-uniform sampling \citep{tu2013dynamic}, account for nonlinear observables to perform Koopman analyses for both deterministic and stochastic datasets \citep{williams2015data}, and improve its robustness to noise using L+S decomposition \citep{wynn2013optimal, schmid2022dynamic} or forward-backward mapping \citep{dawson2016characterizing}. Furthermore, the processing of large datasets can be facilitated through parallelisation of QR-decomposition \citep{sayadi2016parallel}. By partitioning data sequences based on frequency, DMD also leads to improved resolution of flow features \citep{kutz2016multiresolution}. DMD has also inspired other data-driven perturbation analysis methods. For example, \citet{herrmann2021data} introduced an approach to perform resolvent analyses and demonstrated its efficiency on 3D transitional channel flow. However, all these schemes can only approximate a subset of the spectral features of the linear operator. Currently, there is still no robust data-driven framework to perform SVD of the evolution operator from snapshots, which is essential for transient growth analyses.	
	
	From a broader perspective, the concept underlying DMD and its related data-driven perturbation analysis techniques can be understood as a form of Galerkin projection \citep{williams2015data, colbrook2023multiverse}. For example, standard DMD projects the linear operator onto proper orthogonal decomposition (POD) modes derived from snapshots, while optimal mode decomposition (OMD) \citep{wynn2013optimal} projects the same operator onto an optimal low-order subspace determined by a conjugate gradient algorithm. Similarly, data-driven resolvent analysis projects the resolvent operator onto approximated eigenmodes obtained through DMD. These projection methods are used to construct low-dimensional models, where the choice of the subspace significantly influences the accuracy. This is evident from the study in data-driven resolvent analysis, which shows that the resolvent modes extracted from trajectories initialised with random disturbances, optimal forcing, and localised impulses can vary considerably, underscoring the sensitivity of the results to the subspace selection \citep{herrmann2021data}. 
	
	In the Galerkin framework, POD is often utilised to construct these basis sets since it retains the modes biased toward large, high-energy scale \citep{girfoglio2021pod} and the initialisation of trajectories for extracting POD modes often depends on the specific problem context within POD-Galerkin frameworks \citep{hijazi2020data, ballarin2016pod, ballarin2015supremizer}. In this study, we utilise both one-dimensional (1D) and two-dimensional (2D) Hermite polynomials for the initialisation in POD modes extraction, as they serve as the orthogonal basis for Cauchy problems \citep{williams2015data}. {Hermite polynomials, originally introduced by Charles Hermite, play a key role in solving the Hermite differential equation and have broad applications in mathematics and physics \citep{aboites2019simple}. The 1D Hermite polynomials describe the eigenfunctions of the quantum harmonic oscillator \citep{deniz2019quantum}. When extended to 2D, Hermite polynomials are widely adopted in the study of quantum and optical systems to analyse and model complex wave patterns \citep{escalante2008multi,van1996feature,marisa2015bufferless}. Their orthogonality and functional characteristics are expected to enhance the efficiency of capturing and representing key modes in complex systems. It should be noted that the choice of Hermite polynomials is not unique. Other families of orthogonal polynomials (e.g., Chebyshev or Legendre) could be used. The key requirement is to provide spatially diverse, linearly independent inputs, as the subsequent POD modes define the effective low-dimensional basis for the analysis.}
	
	In this work, we present an algorithm designed to extract transient energy growth and associated optimal initial perturbations and their outcomes directly from snapshots of perturbations. This method leverages POD to project the evolution operator onto a reduced-dimensional space. We demonstrate that by selecting appropriate initial conditions, e.g. in the form of Hermite polynomials, the POD modes of their trajectories are sufficient to incorporate optimal initial perturbations. This enables transient growth analysis through direct SVD of the evolution operator in a lower-dimensional space, eliminating the need for deriving and solving adjoint equations.
	
	The structure of this paper is as follows. In \cref{sec: Operator-based method}, we review the model-based formulation of transient growth analyses, then give a description of the proposed data-driven method in \cref{sec: Data-driven method} and demonstrate its effectiveness through three examples and discuss the limitations and potential applications in \cref{sec: Examples}. Finally, the conclusions are presented in \cref{sec: Summary}. 
	
	\section{Model-based transient growth}\label{sec: Operator-based method}
	Before delineating the data-driven transient growth analysis, we first briefly present the algorithms of model-based transient growth.
	
	\subsection{Definition of transient growth}
	For a general description of transient problems, we consider a linear dynamical system in the spatially discretised form:
	\begin{equation}
		\partial_t \boldsymbol{u} = \boldsymbol{A} \boldsymbol{u}, 
		\label{DynamicEquation}
	\end{equation}
	where $\boldsymbol{u} \in \mathbb{C}^N$ denotes the state vector with $N$ representing the dimensionality of the discretised state vectors, and $\boldsymbol{A} \in \mathbb{C}^{N \times N}$ is the linear dynamics operator. Specifically, the state $\boldsymbol{u}$ represents the deviation from the base, which can be the steady state of a laminar flow or the temporal mean of a statistically stationary unsteady flow. The elements of $\boldsymbol{A}$ are typically derived from the underlying physical model, such as the linearised Navier–Stokes (LNS) equations in fluid dynamics. According to (\ref{DynamicEquation}), there is a linear evolution operator that projects the initial condition $\boldsymbol{u}_0$ to the solution $\boldsymbol{u}_{\tau}$:
	\begin{equation}
		\boldsymbol{u}_{\tau} = \mathcal{M}(\tau) \boldsymbol{u}_0,
		\label{EvolutionOperator}
	\end{equation}
	where $\mathcal{M}(\tau)= e^{\boldsymbol{A} \tau} \in \mathbb{C}^{N \times N}$ if $\boldsymbol{A}$ is time-independent. 
	
	To facilitate the presentation, we introduce the following inner product defined on the spatial domain: 
	\begin{equation}
		(\boldsymbol{a}, \boldsymbol{b}) = \boldsymbol{a}^{\mathrm{T}} \boldsymbol{Q} \boldsymbol{b},
		\label{DiscreteInner}
	\end{equation}
	{where $\boldsymbol{a}$ and $\boldsymbol{b}$ are arbitrary vectors in the same inner product space.} The superscript ``$\mathrm{T}$" denotes the complex conjugate, and the positive-definite matrix $\boldsymbol{Q}$ accounts for the grid quadrature \citep{herrmann2021data}. Then the energy of the perturbation at time $\tau$ can be quantified using the $\boldsymbol{Q}$-norm of $\boldsymbol{u}_{\tau}$:
	\[
	E(\tau)=(\boldsymbol{u}_{\tau}, \boldsymbol{u}_{\tau})=\left\|\boldsymbol{u}_{\tau}\right\|_{\boldsymbol{Q}}^{2}. 
	\]
	The $\boldsymbol{Q}$-norm can be related to the standard $L_2$-norm via $\left\|\boldsymbol{u}\right\|_{\boldsymbol{Q}}^{2} = \left\|\boldsymbol{Fu}\right\|_{{2}}^{2}$ with $\boldsymbol{F}$ derived from the Cholesky decomposition $\boldsymbol{Q} = \boldsymbol{F}^{\mathrm{T}}\boldsymbol{F}$ \citep{reddy1993energy}.
	
	The aim of transient growth analyses is to quantify the maximum energy amplification from a given set of initial conditions of the state over a specified time horizon $\tau$. The optimal energy amplification can be expressed as
	\begin{equation}
		G(\tau) = \max_{\boldsymbol{u}_0} \frac{E(\tau)}{E(0)} = \max_{\boldsymbol{u}_0} \frac{\left\|\mathcal{M}(\tau)\boldsymbol{u}_{0}\right\|_{\boldsymbol{Q}}^{2}}{\left\|\boldsymbol{u}_0\right\|_{\boldsymbol{Q}}^{2}} = \sigma_{\rm max} (\boldsymbol{F}\mathcal{M}(\tau)\boldsymbol{F}^{-1}),
		\label{EnergyGrowth}
	\end{equation}
	{where $\sigma_{\rm max}(\cdot)$ denotes the largest singular value of a matrix.} Transient growth analysis can be performed using adjoint methods, including eigenvalue- and optimisation-based techniques. \citet{mao2013calculation} have demonstrated the equivalence of these two approaches for both optimal initial and boundary perturbation problems, and therefore we solely focus on the eigenvalue approach in this work.
	
	\subsection{Eigenvalue approach}\label{sec: Eigenvalue}
	Considering \eqref{EnergyGrowth}, when $\mathcal{M}(\tau)$ is available in a discretised matrix form, transient growth analyses can be conducted using the SVD of the weighted operator:
	\begin{equation}
		\boldsymbol{F}\mathcal{M}(\tau)\boldsymbol{F}^{-1} \boldsymbol{\mit{\Psi}} = \boldsymbol{\mit{\Phi}}\boldsymbol{\mit{\Sigma}}
		\label{SVD},
	\end{equation}
	where $\boldsymbol{\mit{\Sigma}}= \mathrm{diag}({\sigma}_1, {\sigma}_2, \ldots, {\sigma}_{N}) \in \mathbb{R}^{N \times N}$ contains real and non-negative singular values and ${\sigma}_1 \geq {\sigma}_2 \geq, \ldots,\geq {\sigma}_{N} \geq 0$. The matrices  $\boldsymbol{\mit{\Phi}}=[\boldsymbol{\mit{\phi}}_1, \boldsymbol{\mit{\phi}}_2, \ldots, \boldsymbol{\mit{\phi}}_{N}] \in \mathbb{C}^{N \times N}$ and $\boldsymbol{\mit{\Psi}}=[\boldsymbol{\mit{\psi}}_1, \boldsymbol{\mit{\psi}}_2, \ldots, \boldsymbol{\mit{\psi}}_{N}] \in \mathbb{C}^{N \times N}$ contain the left and right singular vectors, respectively. Clearly, the largest singular value ${\sigma}_1$ corresponds to the square root of the maximum energy amplification, while the associated right and left singular vectors, $\boldsymbol{\mit{\psi}}_1$ and $\boldsymbol{\mit{\phi}}_1$, represent the optimal initial perturbation and its resulting optimal outcome, respectively.
	
	In practice, however, the explicit matrix form of $\mathcal{M}(\tau)$ is unavailable.  \citet{barkley2008direct} introduced a matrix-free method for transient growth analyses. {As defined by \citet{courant2008methods}, there exists an adjoint operator  $\boldsymbol{L}^*$ for any linear operator $\boldsymbol{L}$ such that
		\begin{equation}
			(\boldsymbol{b}, \boldsymbol{L} \boldsymbol{a}) = (\boldsymbol{L}^* \boldsymbol{b}, \boldsymbol{a}).
			\label{ADJOINT}
	\end{equation}}
	Then the energy amplification in (\ref{EnergyGrowth}) can be expressed as
	\begin{equation}
		G(\tau )=\underset{{\boldsymbol{u}_{0}}}{\mathop{\max }}\,\frac{\left( {\boldsymbol{u}_{0}},\mathcal{M}^*(\tau)\mathcal{M}(\tau){\boldsymbol{u}_{0}} \right)}{\left( {\boldsymbol{u}_{0}},{\boldsymbol{u}_{0}} \right)},
		\label{Gain_adjoint}
	\end{equation}
	where $\mathcal{M}^*(\tau) \in \mathbb{C}^{N \times N}$ is the adjoint of $\mathcal{M}(\tau)$ based on (\ref{ADJOINT}). The relationship between $\mathcal{M}^*(\tau)$ and $\mathcal{M}^{\mathrm{T}}(\tau)$ is given by
	\[
	\mathcal{M}^*(\tau) =\boldsymbol{Q}^{-1} \mathcal{M}^{\mathrm{T}}(\tau) \boldsymbol{Q}.
	\]
	
	The eigenvalue approach involves projecting the joint operator $\mathcal{M}^*(\tau)\mathcal{M}(\tau)$ onto a low-dimensional space spanned by the Krylov sequence, enabling direct computation of the leading eigenvalues and eigenvectors using the Arnoldi method:
	\[
	\mathcal{M}^*(\tau)\mathcal{M}(\tau)\boldsymbol{\mit{\Psi}}_J = \boldsymbol{\mit{\Sigma}}^2_J \boldsymbol{\mit{\Psi}}_J, 
	\]
	where $\boldsymbol{\mit{\Sigma}}^2_J = \mathrm{diag}({\sigma}^2_1, {\sigma}^2_2, \ldots, {\sigma}^2_{J}) \in \mathbb{R}^{J \times J}$ contains $J$ leading eigenvalues of the joint operator, with $J$ being the length of the Krylov sequence, and $\boldsymbol{\mit{\Psi}}_J = [\boldsymbol{\mit{\psi}}_1, \boldsymbol{\mit{\psi}}_2, \ldots, \boldsymbol{\mit{\psi}}_{J}] \in \mathbb{C}^{N \times J}$ is the matrix of corresponding Ritz eigenvectors. This process requires $J-1$ integrations of the joint operator, i.e. the linearised operator and its adjoint. Since the matrix $\mathcal{M}^*(\tau)\mathcal{M}(\tau)$ is symmetric and self-adjoint, its eigenvalues are real, and the maximum energy amplification $G(\tau)$ corresponds to the largest eigenvalue of this joint operator matrix. Here, the matrix  $\boldsymbol{\mit{\Psi}}_J$ corresponds to the right singular vectors defined in \eqref{SVD}, while the left singular vectors in $\boldsymbol{\mit{\Phi}}_J$ can also be obtained by post-processing computation.
	
	Clearly, the process above relies on the adjoint equations. To circumvent this limitation, we project the forward operator onto a lower-dimensional subspace and subsequently bypass the derivation and calculation of the adjoint equations.

	\section{Data-driven optimal growth analysis}\label{sec: Data-driven method}
	In this section, we outline the algorithm for data-based transient growth calculation, where the initial dataset and low-dimensional space construction are essential. 
	
	\subsection{Initial matrix construction}\label{sec: X construction}
	This data-driven method employs the Galerkin projection to approximate the operator in a lower-dimensional space, a process that will be further elaborated in \cref{LDP}. In this Galerkin framework, POD is typically used to construct basis functions, as it efficiently captures a low-dimensional representation of the dominant spatio-temporal structures in dynamical systems \citep{sirovich1987turbulence,berkooz1993proper,lumey2012stochastic}. 
	
	Supposing a set of initial conditions has been selected, we first capture $K$ flow trajectories which are solutions to the linearised dynamical equations. {The snapshots of the $k$th trajectory initialised by the $k$th element of a selected set (such as Hermite polynomials or their variations) $\boldsymbol{q}^k_{t_0}$, where $k \in \{1, 2, \ldots, K\}$, are assembled into the trajectory matrix $\boldsymbol{D}^k_{1:S} \in \mathbb{C}^{N \times S}$: 
		\begin{equation}
			\boldsymbol{D}^k_{1:S} = [\boldsymbol{q}^k_{t_1}, \boldsymbol{q}^k_{t_2}, \boldsymbol{q}^k_{t_3}, \ldots, \boldsymbol{q}^k_{t_{S}}],
			\label{trajec}
		\end{equation}
		where the subscripts ${t_1, t_2, \ldots, t_{S}}$ denote discrete time points. To ensure satisfaction of the continuity constraint in fluid flow examples, $\boldsymbol{q}^k_{t_0}$ is typically excluded from $\boldsymbol{D}^k_{1:S}$. In other words, the perturbed field is first integrated forward to $t_1$ using the linear solver, and this short-time integration step guarantees that the subsequent velocity field satisfies the continuity equation to numerical accuracy.} It is suggested that $t_{S} \geq \tau$, ensuring that the data contains sufficient transient information at time $\tau$ {where the transient mechanism is of interest.} Then, a rank-$R$ truncated SVD of $\boldsymbol{D}^k_{1:S}$ yields the matrix $[\boldsymbol{u}^{(k,1)}, \boldsymbol{u}^{(k,2)}, \boldsymbol{u}^{(k,3)}, \ldots, \boldsymbol{u}^{(k,R)}]$, containing the first $R$ left-singular vectors of $\boldsymbol{D}^k_{1:S}$, where $R \leq S \ll N$. These singular vectors are then assembled as columns of $\tilde{\boldsymbol{X}}$ in the form of
	\begin{equation}
		\tilde{\boldsymbol{X}}=\left[\boldsymbol{u}^{(1,1)}_0, \boldsymbol{u}^{(1,2)}_0, \ldots, \boldsymbol{u}^{(1,R)}_0 \mid \\
		\boldsymbol{u}^{(2,1)}_0, \boldsymbol{u}^{(2,2)}_0, \ldots, \boldsymbol{u}^{(2,R)}_0 \mid \\ \dots \mid \\ 	\boldsymbol{u}^{(K,1)}_0, \boldsymbol{u}^{(K,2)}_0, \ldots, \boldsymbol{u}^{(K,R)}_0\right],
		\label{Xmatrix}
	\end{equation}
	where $\tilde{\boldsymbol{X}}$ is of size $N \times (K \times R)$ and the subscript ``$0$" is added to highlight that they will be used as a pool of initial conditions in the following process. 
	
	In practice, the selection of $K$ or $R$ may vary considerably, depending on the number of POD modes required to accurately synthesise the optimal modes for the problem. {Specifically, the rank truncation parameter $R$ is determined using a cumulative energy criterion based on the singular values of $\boldsymbol{D}^k_{1:S}$. In this study, $R$ is chosen such that the retained modes capture at least 99\% of the total energy, which has been found to be sufficient to accurately represent the dominant transient dynamics while filtering out low-energy noise.} In case studies, a fixed value of $R$ is selected based on the singular values of $\boldsymbol{D}^1_{1:S}$, and the size of $\tilde{\boldsymbol{X}}$ is increased by expanding $K$ to improve convergence while retaining the same number of SVD modes for $\boldsymbol{D}^2_{1:S}, \boldsymbol{D}^3_{1:S}, \dots, \boldsymbol{D}^K_{1:S}$. {As a practical rule of thumb, as $K$ or $R$ increases, the additional vectors in the reduced subspace $\mathbb{V}$, spanned by the SVD modes of $\boldsymbol{X}$, tend to become linearly dependent if the dynamical system is low-rank. This saturation of the subspace can serve as a practical guideline for selecting appropriate values of $K$ or $R$.} {Other modal decomposition techniques, e.g. the spectral POD (SPOD) \citep{nekkanti2021frequency}, balanced POD (BPOD) \citep{flinois2015projection}, or even convolutional neural networks \citep{murata2020nonlinear}, in addition to snapshot POD, may be used here to reduce the memory requirements of the proposed data-driven algorithm and potentially extend it to more complex scenarios. However, while these methods can increase computational efficiency in subsequent steps of \cref{LDP}, they introduce additional computational costs during the initial matrix construction process or are not completely adjoint-free.}
	
	To effectively capture transient information, Hermite polynomials are introduced to enhance the POD mode generation process in the selection of $\boldsymbol{q}^k_{t_0}$. The details of both 1D and 2D forms of Hermite polynomials are given in Appendix \ref{HPolynomials}. The POD modes, derived from trajectories which are initialised with Hermite polynomials or their variants, are demonstrated to construct accurate reduced-order models. Here, using the Hermite polynomials as the initial guess significantly improves the efficiency of the scheme, as will be compared against random noise in \cref{sec: SFBFS}.
	
	\subsection{Data-driven algorithm}\label{LDP}
	The proposed data-driven transient growth analysis relies on the Galerkin projection of the governing operator onto a space spanned by orthogonal basis vectors, allowing the system state to be approximated as a combination of these vectors. For the convergence test, we establish an initial matrix $\boldsymbol{X}$, composed of the first $n$ column vectors of $\tilde{\boldsymbol{X}}$, such that $\tilde{\boldsymbol{X}} \supseteq \boldsymbol{X} \in \mathbb{C}^{N \times n}$, with $1 \leq n \leq (K \times R)$ and $n \ll N$. To facilitate the process of matrix manipulation and low-dimensional space construction, we perform a full-rank SVD of $\boldsymbol{X}$: 
	\begin{equation}
		\boldsymbol{X}=\boldsymbol{\mit{V}} \boldsymbol{\mit{\Lambda}}\boldsymbol{W}^{\mathrm{T}},
		\label{SVDX}
	\end{equation}
	where $\boldsymbol{\mit{V}} = [\boldsymbol{v}_1, \boldsymbol{v}_2, \ldots, \boldsymbol{v}_n] \in \mathbb{C}^{N \times n}$ is a unitary matrix and spans a reduced basis space $ \mathbb{V} = span{\{\boldsymbol{v}_i\}^n_{i=1}} $. $\boldsymbol{\mit{\Lambda}} \in \mathbb{R}^{n \times n}$ is a diagonal matrix of singular values and $\boldsymbol{W} \in \mathbb{C}^{N \times n}$ is the matrix of right singular vectors.
	
	By evolving the column vectors of $\boldsymbol{X}$ forward in time from $t=0 $ to $ t=\tau$ using equation (\ref{DynamicEquation}), where the transient growth analysis is to be performed, aligning with equation (\ref{EvolutionOperator}), we obtain
	\begin{equation}
		\boldsymbol{Y_{\tau}} = \mathcal{M}(\tau)\boldsymbol{X},
		\label{EVOLVING}
	\end{equation}
	where $\boldsymbol{Y_{\tau}} \in \mathbb{C}^{N \times n}$ is called the outcome matrix here. For clarity in the subsequent derivation, we use $\boldsymbol{u}_{0}$ to denote a column vector from $\boldsymbol{X}$, and $\boldsymbol{u}_{\tau}$ to denote the corresponding column in $\boldsymbol{Y_{\tau}}$. It should be noted that we can collect the snapshots at an arbitrary time instance $t \in [0,\tau]$ during a single forward integration, enabling the computation of transient growth over multiple time instances. This contrasts with model-based methods, which require renewed forward and backward integration for each time instance.
	
	Assuming that $\boldsymbol{Y_{\tau}}$ can be expressed as linear combinations of columns of $\boldsymbol{\mit{V}}$, which is reasonable given that $\boldsymbol{X}$ is constructed by POD modes from transient trajectories as described in \cref{sec: X construction}, then the final and initial states can be approximated in the space $\mathbb{V}$ as:  
	\begin{equation}
		\boldsymbol{u}_{\tau}=\boldsymbol{V} \hat{\boldsymbol{u}}_{\tau}, \quad \boldsymbol{u}_{0}=\boldsymbol{V} \hat{\boldsymbol{u}}_{0},
		\label{Galerkin}
	\end{equation}
	with $\hat{\boldsymbol{u}}_{\tau}$ and $\hat{\boldsymbol{u}}_0$ being the reduced states of size $n \times 1$. 
	
	Considering the projection in (\ref{Galerkin}), we can then construct a reduced order forward operator $\hat{\mathcal{M}}(\tau) = \boldsymbol{V}^{\mathrm{T}} {\mathcal{M}}(\tau) \boldsymbol{V}$, which satisfies:
	\[
	\hat{\boldsymbol{u}}_{\tau}=\hat{\mathcal{M}}(\tau)\hat{\boldsymbol{u}}_{0},
	\]
	By substituting the SVD in (\ref{SVDX}) into (\ref{EVOLVING}), we can approximate $\hat{\mathcal{M}}(\tau)$ as follows:
	\[
	\hat{\mathcal{M}}(\tau) = \boldsymbol{V}^{\mathrm{T}} \boldsymbol{Y_{\tau}}\boldsymbol{W}\boldsymbol{\mit{\Lambda}}^{-1}.
	\]
	Since the energy is defined in the discrete form, in this reduced subspace, we need to adjust the inner product to retain the physical meaning in the $L_2$-norm framework:
	\[
	\left\|\boldsymbol{u}\right\|_{\boldsymbol{Q}}^{2}=\hat{\boldsymbol{u}}^{\mathrm{T}} \boldsymbol{V}^{\mathrm{T}} \boldsymbol{Q} \boldsymbol{V} \hat{\boldsymbol{u}}=\left\|\hat{\boldsymbol{F}} \hat{\boldsymbol{u}}\right\|_{2}^2,
	\]
	where the matrix $\hat{\boldsymbol{F}} \in \mathbb{C}^{n \times n}$ can be derived from the Cholesky factorization of $\boldsymbol{V}^{\mathrm{T}} \boldsymbol{Q} \boldsymbol{V} = \hat{\boldsymbol{F}}^{\mathrm{T}} \hat{\boldsymbol{F}}$. Thus, the optimal energy growth can be expressed as 
	\[
	G(\tau) = \max_{\boldsymbol{u}_0} \frac{\left\|\boldsymbol{u}_{\tau}\right\|_{\boldsymbol{Q}}^{2}}{\left\|\boldsymbol{u}_0\right\|_{\boldsymbol{Q}}^{2}} = \max_{\boldsymbol{u}_0} \frac{\left\| \hat{\boldsymbol{F}} \hat{\boldsymbol{u}}_{\tau}\right\|_{2}^{2}}{\left\| \hat{\boldsymbol{F}} \hat{\boldsymbol{u}}_0\right\|_{2}^{2}} = \max_{\boldsymbol{u}_0} \frac{\left\| \hat{\boldsymbol{F}} \hat{\mathcal{M}}(\tau) \hat{\boldsymbol{u}}_{0} \right\|_{2}^{2}}{\left\| \hat{\boldsymbol{F}} \hat{\boldsymbol{u}}_0\right\|_{2}^{2}} = \sigma_{\rm max} ( \hat{\boldsymbol{F}} \hat{\mathcal{M}}(\tau) \hat{\boldsymbol{F}}^{-1}).
	\]
	
	Then a standard SVD can be performed to obtain the modes and energy growth in the $n$-dimensional vector space $\mathbb{V}$:
	\[
	\hat{\boldsymbol{F}} \hat{\mathcal{M}}(\tau) \hat{\boldsymbol{F}}^{-1} \hat{\boldsymbol{\mit{\psi}}}_1 = \hat{{\mit{\sigma}}}_1 \hat{\boldsymbol{\mit{\phi}}}_1,
	\]
	where the largest singular value $\hat{\mit{\sigma}}_1$ gives the square root of the transient energy growth at time $\tau$ and the modes in original coordinates as (\ref{SVD}) are represented by:
	\[
	{{\boldsymbol{\mit{\psi}}}_1 = \boldsymbol{V} \hat{\boldsymbol{F}}^{-1} \hat{\boldsymbol{\mit{\psi}}}_1, \quad \boldsymbol{\mit{\phi}}}_1 = \boldsymbol{V} \hat{\boldsymbol{F}}^{-1} \hat{\boldsymbol{\mit{\phi}}}_1.
	\]
	
	The $\boldsymbol{Q}$-norm error ${e}_{\rm mode}$ is defined to quantitatively evaluate the performance of this data-driven method in extracting optimal modes and the relative error ${e}_{\rm growth}$ characterises the error in optimal energy growth calculations, as follows:
	\[
	{e}_{\rm mode} = \frac{\left\|\boldsymbol{\zeta} - {\boldsymbol{\zeta}}_{\mathrm{d}} \right\|_{\boldsymbol{Q}}^{2}}{\left\|\boldsymbol{\zeta}\right\|_{\boldsymbol{Q}}^{2}}, \quad 
	{e}_{\rm growth} = \frac{|{\varepsilon} - {{\varepsilon}}_{\mathrm{d}}|} {{\varepsilon}},
	\]
	where $\varepsilon$ denotes the energy growth and $\boldsymbol{\zeta}$ represents the optimal initial perturbation or outcome, both serve as the reference from the model-based method here, while the subscript ``$\mathrm{d}$" refers to results obtained using the data-driven method. {Convergence with respect to parameters, particularly the number of initial conditions $n$, can then be assessed by comparing the predicted transient growth and associated modes with reference model-based results. The criterion adopted here is that ${e}_{\rm growth}$ and ${e}_{\rm mode}$ both fall below $10^{-2}$.}
	
	\section{Results}\label{sec: Examples}
	In this section, we apply the established data-driven transient growth analysis algorithm to three representative examples, i,e., the complex Ginzburg–Landau equation with a single velocity component, the backward-facing step flow with two velocity components featuring the Orr transient mechanism \citep{marquet2012convective}, and the Batchelor vortex with three velocity components demonstrating the anti-lift-up transient mechanism \citep{antkowiak2007vortex,mao2012transient}. These examples, with increasing complexity, are well-established in the context of model-based hydrodynamic stability and transient growth studies whose results agree well with those of the proposed data-driven approach as will be presented. {Moreover, while the proposed method shares similarities with DMD in terms of operator approximation, they differ fundamentally in the objective, construction process, and data requirements. These distinctions will be discussed in detail in \cref{sec: Correlation with DMD}.}
	\subsection {Complex Ginzburg–Landau equation}\label{sec: CCGGLLEE}
	
	\begin{figure}
		\includegraphics[width=1\linewidth]{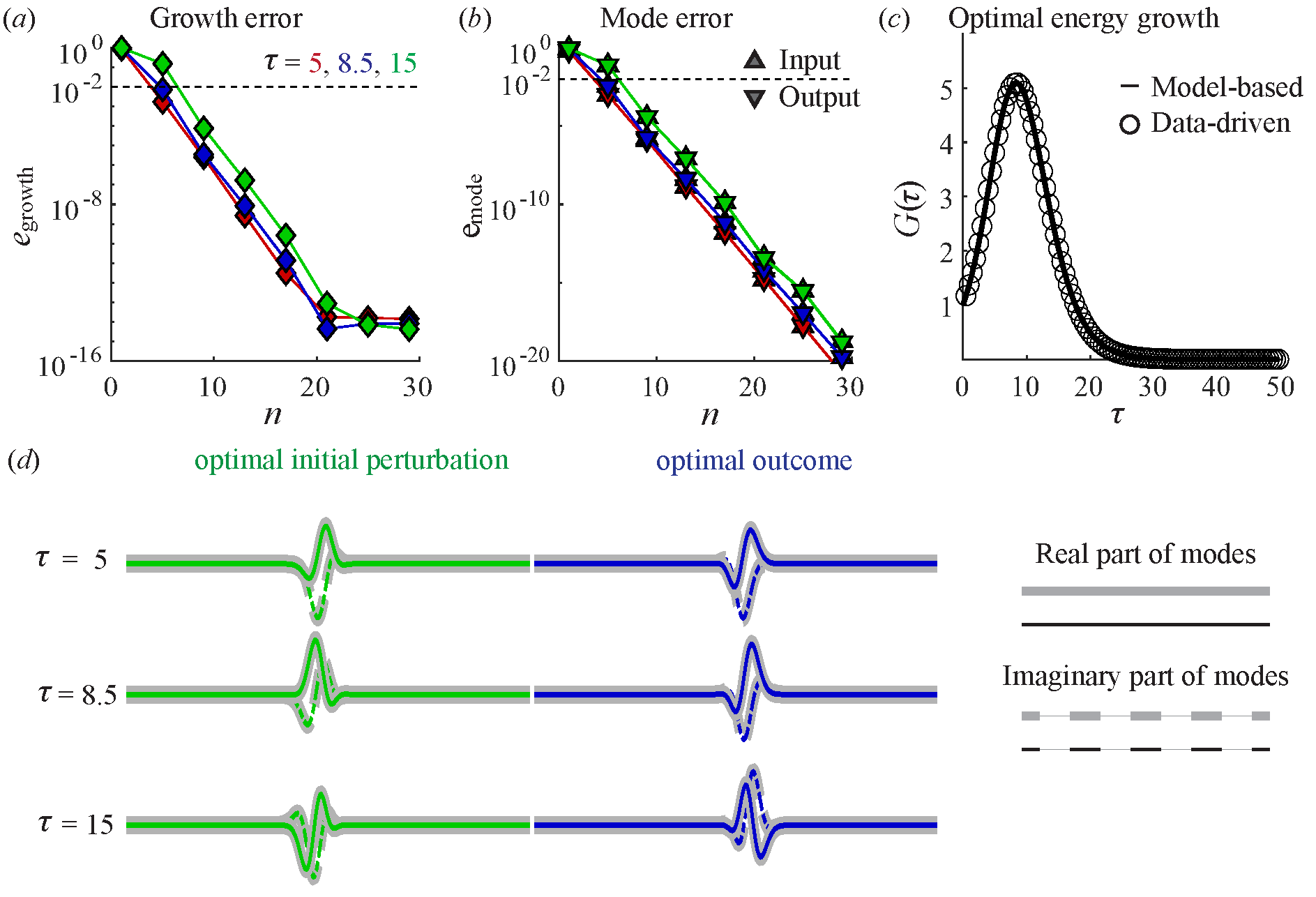}
		\captionsetup{width=1\textwidth}
		\caption{\justifying {Data-driven transient growth analysis of the linearised complex Ginzburg–Landau equation. $(a)$: The relative error between the model-based and the data-driven optimal energy growth as a function of the number of initial conditions in the dataset $\boldsymbol{X}$. $(b)$: The same as $(a)$, but for the $\boldsymbol{Q}$-norm error between the optimal initial perturbations (Input) and outcomes (Output). $(c)$: Optimal energy growth computed by the data-driven method with $n=8$ (circle) and the model-based method (line). $(d)$: Optimal initial perturbations and their corresponding outcomes obtained by the data-driven method with $n=8$ at $\tau=5$, $8.5$, $15$. Solid and dashed lines denote the real and imaginary parts of the modes, respectively, while the grey lines indicate the model-based results for comparison.}}
		\label{fig: 1dcgle}
	\end{figure}
	
	The complex Ginzburg–Landau equation is a widely used model for characterising both the short-term and long-term evolution of small perturbations in spatially developing flows \citep{bagheri2009input} and can be written as
	\[
	\frac{\partial \tilde{u}}{\partial t} = -\eta \frac{\partial \tilde{u}}{\partial x} + \gamma\frac{\partial^2 \tilde{u}}{\partial x^2} + \mu(x)\tilde{u} - c|\tilde{u}|^2 \tilde{u},
	\]
	where the variable $\tilde{u}(x, t)$ represents the complex amplitude, $c$ the complex weight of nonlinearity, $\mu(x)$ the spatial amplification parameter,  $\eta$ the complex advection speed and $\gamma$ the complex diffusion coefficient. By assuming the flow is near the equilibrium state $\tilde{u} = 0$, the linearised governing equation is given by
	\[
	\frac{\partial {u}}{\partial t} = \left(-\eta \frac{\partial}{\partial x}+\gamma\frac{\partial^2}{\partial x^2}+\mu(x)\right) {u},
	\]
	where $u$ denotes small deviations from the equilibrium state, and the corresponding adjoint equation can be represented by
	\[
	-\frac{\partial {u}^*}{\partial t} = \left({\eta}^{\mathrm{T}} \frac{\partial}{\partial x} + {\gamma}^{\mathrm{T}} \frac{\partial^2}{\partial x^2}+{\mu}^{\mathrm{T}}(x)\right){u}^*.
	\]
	The parameters in above equations are selected as $\eta=2+0.4\mathrm{i}, \gamma=1-\mathrm{i}$ and $ \mu(x)= 0.19-0.005{x}^2$, as adopted by \citet{herrmann2021data}, ensuring linearly stable dynamics. The computational domain extends over the range $ -85 < x < 85 $ with homogeneous boundary conditions and is decomposed into $ N = 220 $ uniformly spaced points. In a convergence test, the current resolution results in a relative difference below $ 10^{-6} $ compared with that at $ N = 440 $. The exponential time differencing (ETD) method, as outlined in \citet{cox2002exponential}, is used to evolve both the forward and adjoint equations in the model-based method for comparison.
	
	Following the process outlined in \cref{sec: Data-driven method}, to construct $\tilde{\boldsymbol{X}}$, we employ the first $K = 6$ Gauss-weighted 1D Hermite polynomials to generate trajectories as described in (\ref{trajec}). A total of $S=100$ snapshots, sampled at intervals of $\Delta t=0.5$ time units, are recorded for each trajectory. Subsequently, $R = 5$ POD modes are retained to form $\tilde{\boldsymbol{X}}$, which comprises $30$ column vectors. We then track their evolution to compute both the transient growth and the optimal modes for the time instance $\tau \in \{1,2,\ldots,50\}$.
	
	To assess the convergence performance of the data-driven method, we evaluate the relative error of the optimal energy growth (figure \ref{fig: 1dcgle}$(a)$) and the $\boldsymbol{Q}$-norm error of the modes (figure \ref{fig: 1dcgle}$(b)$) as functions of $n$, i.e. the number of columns in the initial matrix $\boldsymbol{X}$. The results indicate that increasing the number of initial conditions improves the agreement with model-based results. Notably, we can see from these curves that using only $n=8$ initial conditions is sufficient to perform the data-driven algorithm in this case. Furthermore, since snapshots are collected at several time instances during the assembly of the outcome matrix $\boldsymbol{Y}_{50}$, as described in \cref{LDP}, it becomes computationally efficient to compute the envelope of the optimal energy growth, which is in good agreement with that from the model-based method as shown in figure \ref{fig: 1dcgle}$(c)$. 
	
	The optimal initial perturbations and outcomes at $\tau=5, 8.5, 15$, using the data-driven method with $n=8$, are shown in figure \ref{fig: 1dcgle}$(d)$, where we can see that the data-driven method supports successful extraction of transient modes. These results demonstrate that the data-driven transient growth analysis offers an accurate approximation of the transient behaviour over short time intervals in a dynamical system.
	
	\subsection {Flow over a backward-facing step}\label{sec: SFBFS}
	The second example is the backward-facing step flow, where the fluid motion is governed by the incompressible Navier-Stokes (NS) equations:
	\begin{equation}
		\partial_t \tilde{\boldsymbol{u}}=-\tilde{\boldsymbol{u}} \cdot \nabla \tilde{\boldsymbol{u}}-\nabla \tilde{p} + Re^{-1} \nabla^2 \tilde{\boldsymbol{u}}, \quad \nabla \cdot \tilde{\boldsymbol{u}}=0,
		\label{NS}
	\end{equation} 
	with $\tilde{p}$ representing the modified or kinematic pressure and $\tilde{\boldsymbol{u}}$ denoting the velocity vector field. The Reynolds number $Re$ is set to $500$ where the bifurcation to three-dimensional state at $Re=748$ is avoided to ensure an asymptotically stable flow \citep{barkley2002three}. {At the inflow boundary $\partial \Omega_i$, a parabolic velocity profile is prescribed. On the solid walls $\partial \Omega_w$, no-slip conditions are enforced. At the downstream outflow boundary $\partial \Omega_o$, a zero-traction condition is applied, with the pressure set to zero. Collectively, the boundary conditions can be expressed as
		\[
		\begin{aligned}
			\tilde{\boldsymbol{u}}(\partial \Omega_i, t) & =\left(4 y-4 y^2, 0,0\right), \\
			\tilde{\boldsymbol{u}}(\partial \Omega_w, t) & =(0,0,0), \\
			\partial_x \tilde{\boldsymbol{u}}(\partial \Omega_o, t) & =(0,0,0), \quad \tilde{p}(\partial \Omega_o, t)=0.
		\end{aligned}
		\]
	}By decomposing the flow field into a base flow and a perturbation and omitting the interaction of the perturbation with itself, we obtain the LNS equations that govern the evolution of perturbations,
	\begin{equation}
		\partial_t \boldsymbol{u}  =-(\boldsymbol{U} \cdot \nabla) \boldsymbol{u}-\left(\boldsymbol{u} \cdot \nabla\right) \boldsymbol{U}-\nabla p+R e^{-1} \nabla^2 \boldsymbol{u}, \quad \nabla \cdot \boldsymbol{u} =0,
		\label{LNS}
	\end{equation} 
	\begin{figure}
		\includegraphics[width=1\linewidth]{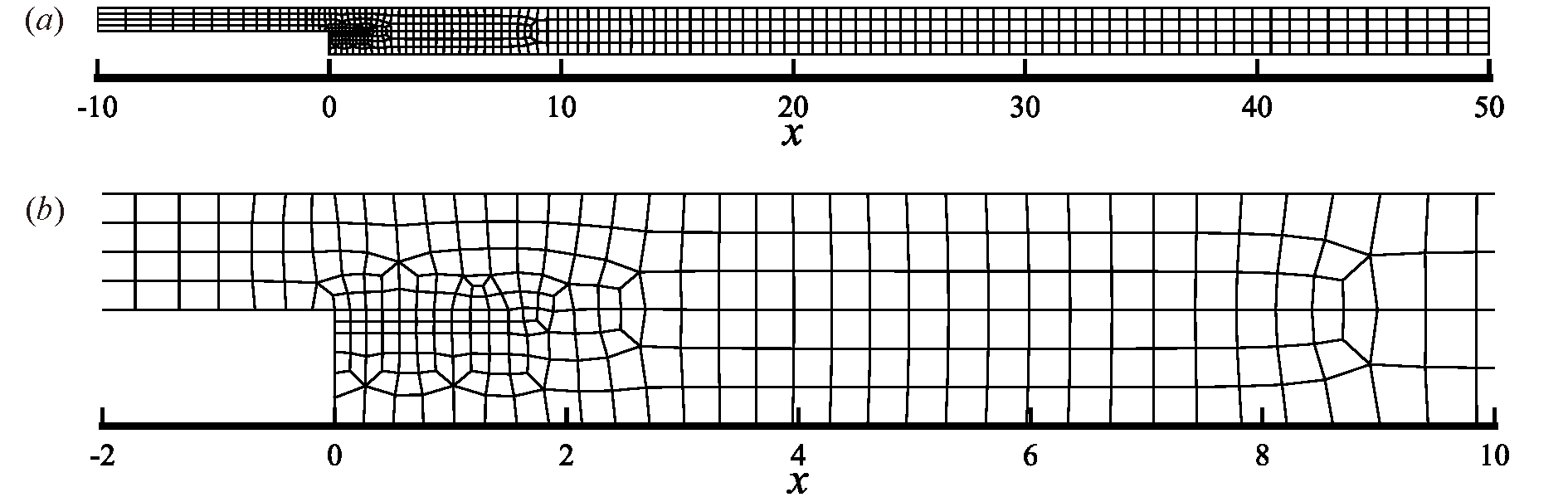}
		\captionsetup{width=1\textwidth}
		\caption{\justifying  Spectral elements in the computational domain: (a) overall domain and (b) domain close to the step.}
		\label{fig: bfsmesh}
	\end{figure}
	where $\boldsymbol{u}$ represents the perturbation of velocity with two components $(u, v)^{\mathrm{T}}$ in streamwise and vertical directions. $\boldsymbol{U}$ is the base flow velocity satisfying the steady NS equations and $p$ is the pressure perturbation. {The adjoint equations, as outlined in \citet{barkley2008direct}, take the form:
		\begin{equation}
			\partial_t \boldsymbol{u}^{*}  =(\boldsymbol{U} \cdot \nabla) \boldsymbol{u}^{*}+\left(\boldsymbol{u}^{*} \cdot \nabla\right) \boldsymbol{U}+\nabla p^{*}-Re^{-1} \nabla^2 \boldsymbol{u}^{*} \quad \text { with } \nabla \cdot \boldsymbol{u}^{*} =0,
			\label{ALNS}
		\end{equation} 
		and are subject to the boundary condition $\boldsymbol{u}^{*}(\partial \Omega, t) =(0,0,0)$ on the entire domain boundary $\partial \Omega$. This condition, in combination with a homogeneous Dirichlet boundary condition for perturbations, i.e., $\boldsymbol{u}(\partial \Omega, t) =(0,0,0)$, ensures consistency between the forward and adjoint systems \citep{barkley2008direct}.} The NS (Eqn. \ref{NS}), LNS (Eqn. \ref{LNS}) and adjoint (Eqn. \ref{ALNS}) equations in the computational domain depicted in figure \ref{fig: bfsmesh} are discretised using the spectral/hp element method \citep{karniadakis2005spectral} and a polynomial order of 6 is used to ensure computational accuracy \citep{mao2015effects}.
	
	To implement the data-driven approach, we start by generating snapshots with $K=240$ trajectories, where the initial conditions are from 2D Hermite polynomials in Cartesian coordinates. Following the formula in (\ref{TDHc}), the first $12$ polynomials are employed with  $C=1$, which defines the non-zero regions. The polynomials are centred along the $x$-direction, extending from $(x,y)=(-9.5,0.5)$ to $(x,y)=(9.5,0.5)$  with a spacing of $1$. To study the short-term energy growth within $\tau \in [0, 100]$ which have been studied in \citet{blackburn2008convective}, each trajectory evolves over a time span of $t_{S}=100$ with interval $\Delta t=0.25$, ensuring that the recorded snapshots capture the key spatial features of the transient dynamics. {For ease of implementation in numerical simulations, only the streamwise velocity component is initially perturbed. The field at time $t_1=0.25$, denoted by $\boldsymbol{q}^k_{t_1}$, satisfies the continuity equation to numerical accuracy, as discussed in \cref{sec: X construction}.}
	
	\begin{figure}
		\includegraphics[width=1\linewidth]{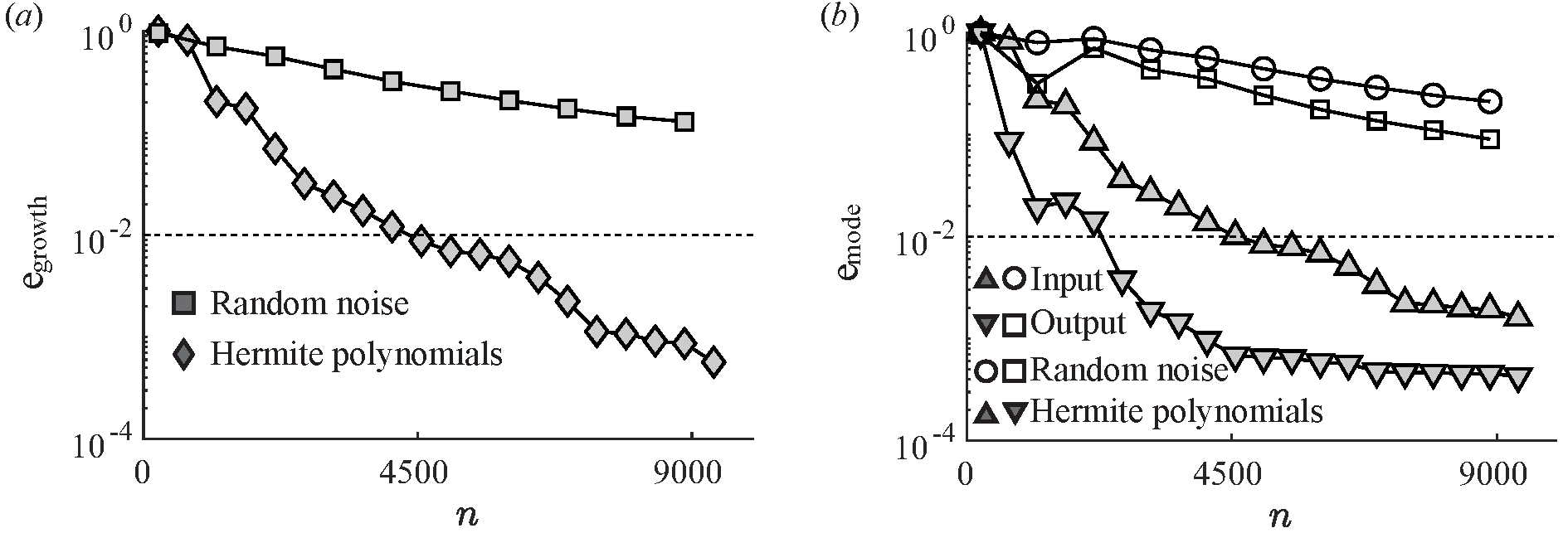}
		\captionsetup{width=1\textwidth}
		\caption{\justifying Comparison of data-driven transient growth analyses of the backward-facing step flow using the dataset obtained from LNS initialised with random noise and Hermite polynomials. $(a)$: Relative error between the model-based and data-driven transient growth at $\tau=60$ as a function of the number of initial conditions in the dataset $\boldsymbol{X}$. $(b)$: Corresponding $\boldsymbol{Q}$-norm error for optimal initial perturbations (Input) and outcomes (Output).}
		\label{fig: 2dbfscurves}
	\end{figure}
	
	\begin{figure}
		\includegraphics[width=1\linewidth]{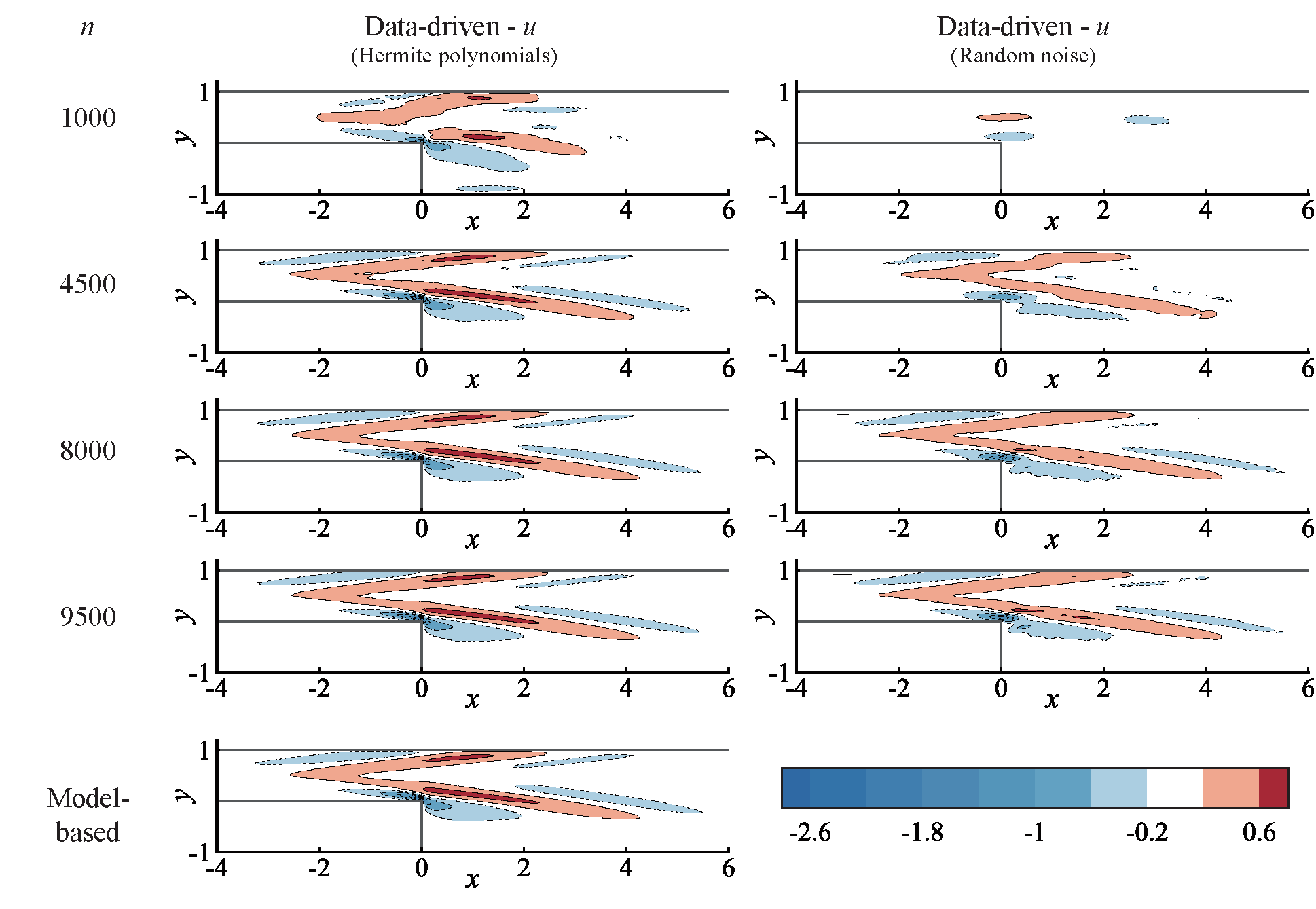}
		\captionsetup{width=1\textwidth}
		\caption{\justifying  {The data-based optimal initial perturbations at $\tau = 60$ from Hermite polynomials and random noise using different numbers of initial conditions in the dataset $\boldsymbol{X}$.}}
		\label{fig: 2dbfsmodestau60}
	\end{figure}
	
	It is crucial to note that the components of the velocity field at each grid point in the $x$ and $y$ directions are concatenated into a single column vector, resulting in $N \textgreater 5 \times 10^4$. To optimise the memory usage, only the first $R=40$ POD modes per trajectory are retained. We then assemble the obtained POD modes into the initial matrix $\boldsymbol{X}$, which is arranged according to the rank of Hermite polynomials. The column vectors in $\boldsymbol{X}$ are then evolved forward in time to generate the outcome matrices $\boldsymbol{Y}_{\tau}$ at specific time instances for transient growth analyses according to \cref{LDP}. For validation purposes, transient growth analysis is performed at $\tau=60$, corresponding to the evolution time of optimal initial perturbations identified in \citet{blackburn2008convective}.
	
	To illustrate the importance of the construction of $\boldsymbol{X}$, {we compare the results using the data generated by Hermite polynomials and random noise, which is zero-mean, spatially uncorrelated, and with a fixed standard deviation applied to the streamwise velocity component at each mesh node}, and present the relative error of energy growth and the $\boldsymbol{Q}$-norm error of the modes in figure \ref{fig: 2dbfscurves}$(a)$ and \ref{fig: 2dbfscurves}$(b)$, respectively. These figures reveal consistent convergence of both energy growth and modes as the amount of data increases. However, compared to Hermite polynomials, the convergence speed of growth and modes using random noise is much slower and the errors of transient growth and modes are around $10 \% $ even with over $n=9000$ initial conditions in $\boldsymbol{X}$, which is about the upper limit of a computer with 64 GB RAM. In contrast, the errors can be reduced to below 1\% using only $n=4500$ initial conditions generated by the data from Hermite polynomials. Given the original forward operator dimensionality, this represents over two orders of magnitude reduction. Besides, the transient information at any time instance in $\tau \in (0,60)$ can be obtained by collecting the corresponding outcome matrix $\boldsymbol{Y}_{\tau}$ as discussed in \cref{LDP}, which can be more efficient than the model-based method when a large number of $\tau$ are considered.
	
	To further illustrate the benefit of employing the Hermite polynomials, the optimal initial perturbations at $\tau = 60$ using the two forms of basis sets are shown in figure \ref{fig: 2dbfsmodestau60}. It can be observed that the data generated from {random noise} is far less efficient in predicting the optimal initial perturbation, particularly around the vicinity of the step edge where the energy is sharply concentrated.
	
	Unlike the complex Ginzburg-Landau equation case, the convergence speed of the optimal initial perturbations is slightly slower than that of the optimal outcomes, which is consistent with observations from the data-driven resolvent analysis \citep{herrmann2021data}. This can be explained by the fact that the spatial support of the POD modes tends to resemble the response modes rather than the forcing modes, which then require more data to converge.
	
	\subsection {Axially decomposed Batchelor vortex}\label{sec: batchelorExample}
\begin{figure}
		\includegraphics[width=1\linewidth]{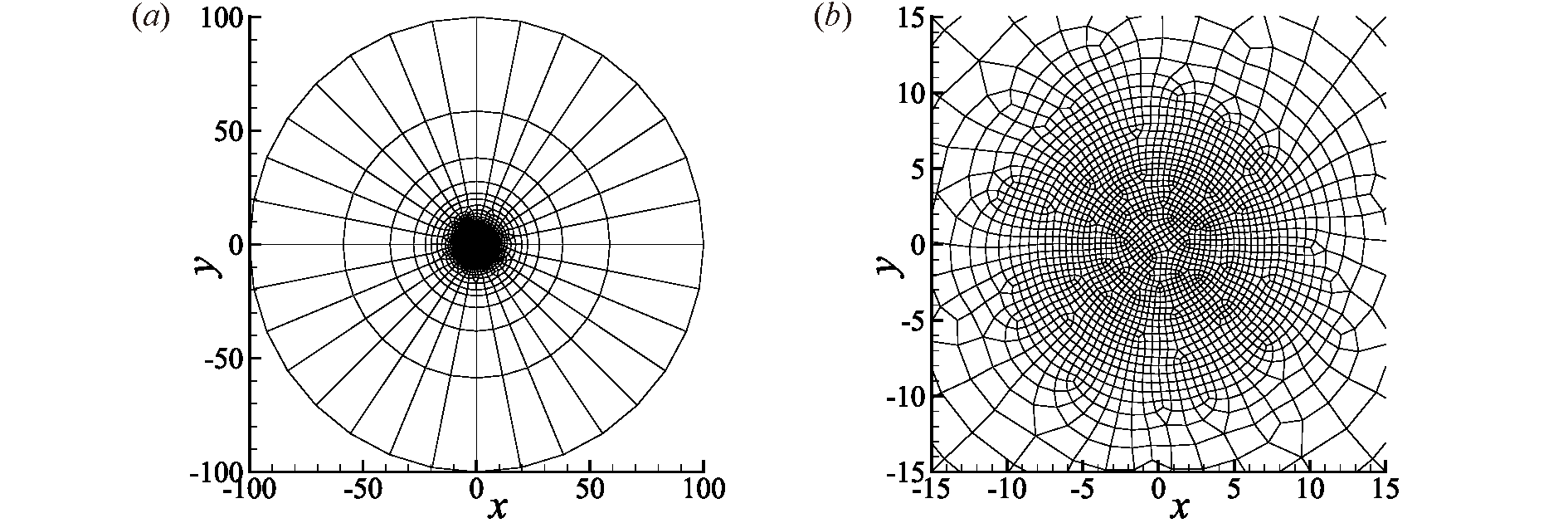}
		\captionsetup{width=1\textwidth}
		\caption{\justifying Computational mesh for transient growth calculation in the axially decomposed global analysis of the Batchelor vortex. $(a)$: complete mesh; $(b)$: mesh around the vortex core.}
		\label{fig: vortexmesh}
	\end{figure}

	To validate the generalisability of the methodology, the Batchelor vortex with three velocity components, which is widely studied in stability and transient growth analyses, is considered. According to \citet{batchelor1964axial}, the non-dimensional base flow velocity $\boldsymbol{U}=(U, V, W)^{\mathrm{T}}$ in cylindrical coordinates $(r,\theta,z)$, where $z$ denotes the axial flow direction, can be described with:
	\[
	\left\{\begin{aligned}
		U & =0, \\
		V & =q\frac{1-e^{-r^2}}{r}, \\
		W & =a + e^{-r^2},
	\end{aligned}\right.
	\]
	where the viscous decay term is neglected to keep the base in a steady form when considering instability or non-normality problems. The perturbations, which are governed by the LNS equations (see Eqn. \ref{LNS}), are decomposed in the axial direction in both model-based and data-driven global analyses. We adopt the parameters consistent with those detailed in \citet{mao2012transient}, where the external non-dimensional free-stream axial velocity is $a=0$ and the swirl strength $q=3$. The Reynolds number is set to $Re=1000$ and the streamwise wave number $0$, which maximises the local transient growth, is adopted. The numerical solver employed is consistent with that in \cref{sec: SFBFS}, and the computational domain and mesh configuration are depicted in figure \ref{fig: vortexmesh}. For convenience, the mesh and subsequent result contours are presented in the Cartesian coordinate system $(x,y,z)$, where $z$ denotes the streamwise direction. 
	\begin{figure}
		\centering
		\includegraphics[width=1\linewidth]{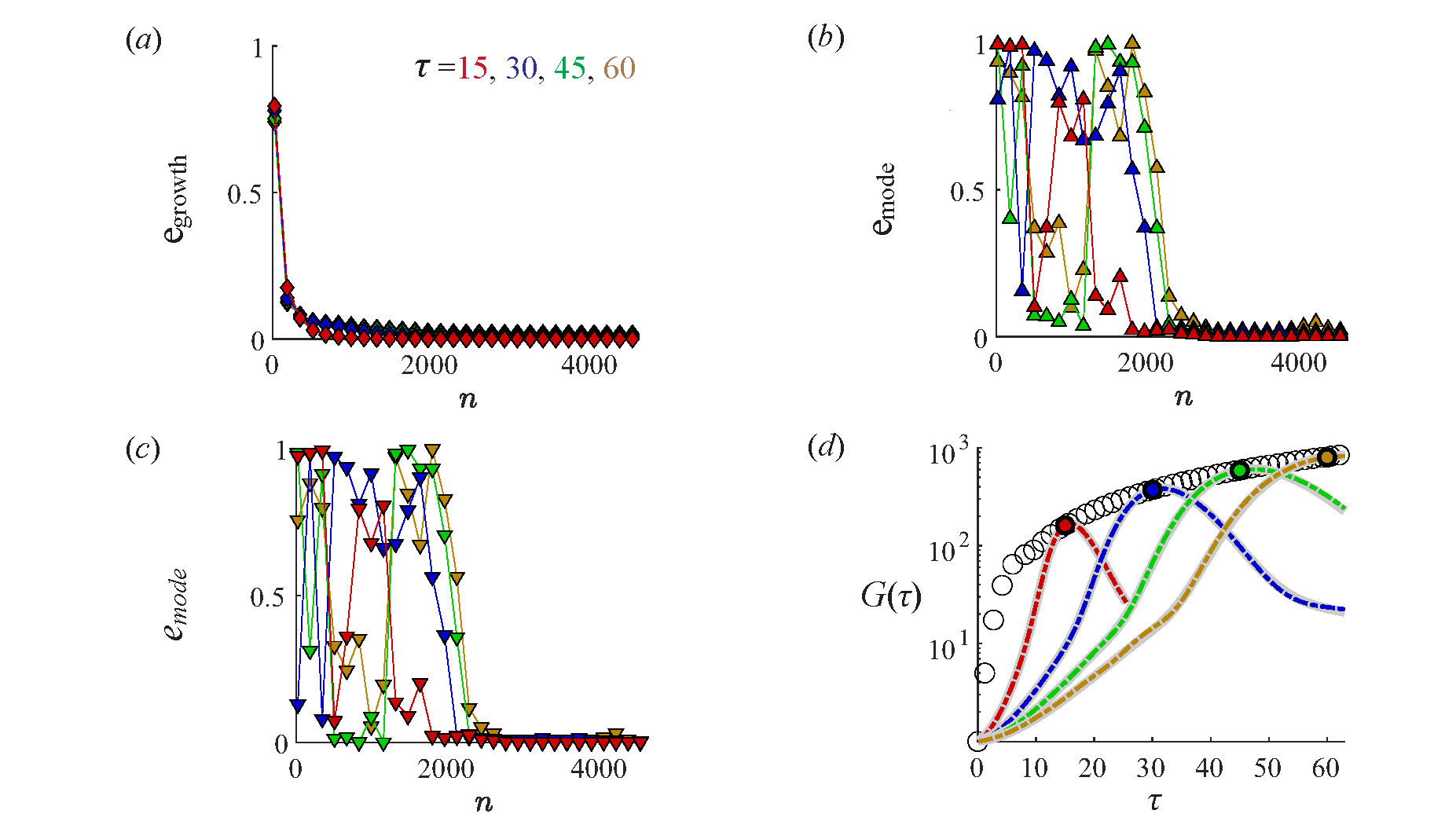}
		\captionsetup{width=1\textwidth}
		\caption{\justifying Data-driven transient growth analysis of the axially decomposed Batchelor vortex at $Re=1000$ and axial wavenumber 0. $(a)$: The relative error between the model-based and the data-driven optimal energy growth at $\tau=15$, $30$, $45$, $60$ as a function of the number of initial conditions in the dataset $\boldsymbol{X}$. $(b)$: The same as $(a)$, but for the $\boldsymbol{Q}$-norm error between the optimal initial conditions. $(c)$: The $\boldsymbol{Q}$-norm error between the optimal outcomes. $(d)$: The envelope of optima (hollow circle) with curves of linear energy evolution starting from the four optimal initial conditions, which are calculated through the data-driven method. The solid circles mark the points at which the curves of linear energy growth are tangential to the envelope.}
		\label{fig: 3dvortexcurves}
	\end{figure}
	\begin{figure}
		\includegraphics[width=1\linewidth]{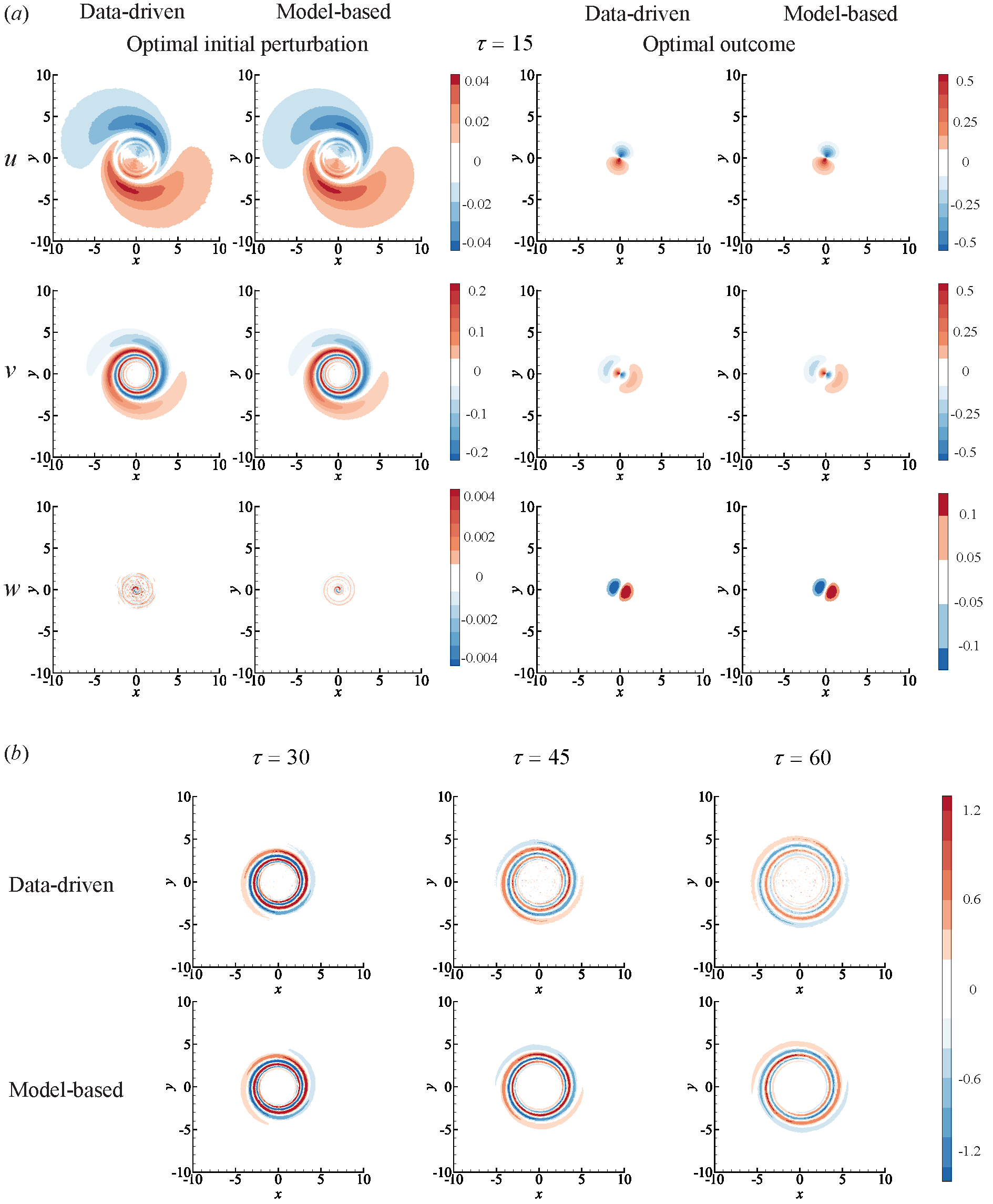}
		\captionsetup{width=1\textwidth}
		\caption{\justifying Comparison of optimal modes computed by data-driven and model-based methods. $(a)$: The optimal initial perturbations and outcomes at $\tau = 15$, the velocity in radial $(u)$, azimuthal $(v)$ and axial $(w)$ directions are shown separately. $(b)$: The streamwise vorticity of optimal initial perturbations at $\tau = 30, 45, 60$.}
		\label{fig: 3dvortextau1530}
	\end{figure}
	
	After discretisation, the dimension of the velocity vector exceeds $2.3\times10^5$, an order of magnitude larger than that of the backward-facing step flow case. Due to this high dimensionality, significantly more data is required to ensure accurate results and to mitigate the curse of dimensionality \citep{bellman1966dynamic}.
	
	One way to minimise the initial conditions required is to pre-determine the azimuthal wavenumber of the optimal initial perturbation since the base flow, in this case, is homogeneous in the axial direction and the optimal mode is with a single axial wavenumber \citep{bale2010transient}. Initially, we select the first $10$ Hermite polynomials in polar coordinates as in (\ref{TDHp}) with the parameter $C \in \{1,2,3, \dots, 20\}$. We generate $m=200$ trajectories initialised with these polynomials by integrating LNS equations forwards, recording 1000 snapshots per trajectory at time intervals of $\Delta t=0.6$. The initial matrix, $\boldsymbol{X}$, is constructed using the first $R=10$ POD modes, resulting in the initial conditions matrix $\boldsymbol{X}$ with $n=2000$. This setup is sufficient to determine the azimuthal wavenumber of the optimal initial perturbation. Once the wavenumber is obtained, the same procedure is applied using Hermite polynomials with various $C$ but a fixed azimuthal wavenumber. {While this reduces the number of initial conditions, it remains computationally inefficient because the polynomials are not strictly spatially orthogonal within the domain, leading to overlapping regions and thus computational redundancy.}
	
	\begin{figure}
		\includegraphics[width=1\linewidth]{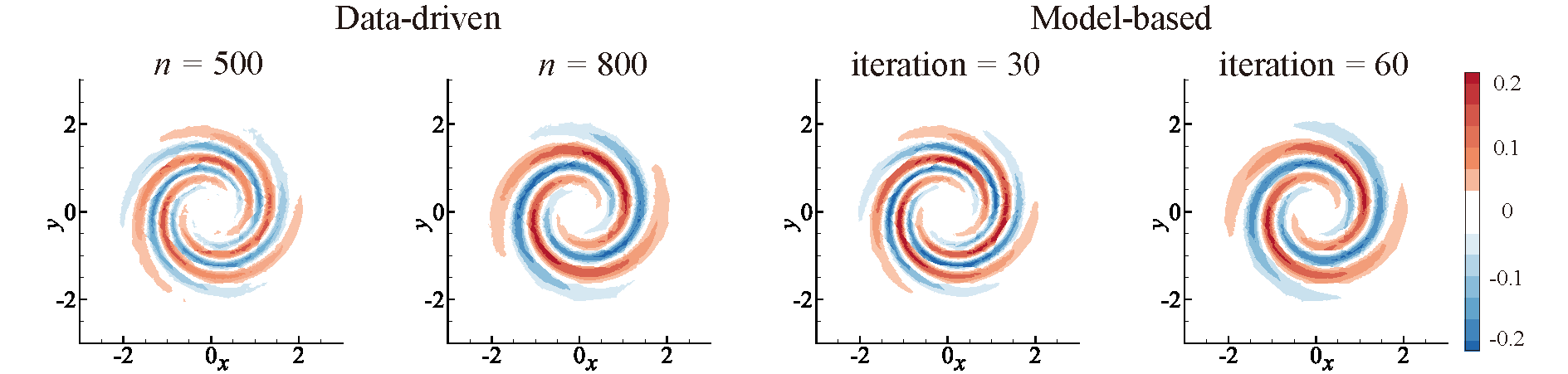}
		\captionsetup{width=1\textwidth}
		\caption{\justifying {Optimal initial condition at $\tau=3$ using the data-driven method with varying data sizes and the model-based method with different iteration counts.}}
		\label{fig: vortexreview}
	\end{figure}
	
	In an alternative approach, we utilise the first $150$ SVD modes of $1200$ polynomials in polar coordinates for the selection of $\boldsymbol{q}^k_{t_0}$ in (\ref{trajec}). These polynomials correspond to the first 6 Hermite polynomials, with the radius parameter $C \in \{0.1, 0.2, 0.3, \dots, 20\}$, where the smaller interval is used to enhance the resolution for capturing features near the vortex core. 
	
	Following the steps in \cref{sec: X construction}, $K=150$ trajectories are then generated and each is recorded with $S=1250$ snapshots with time interval $\Delta t=0.048$. {It should also be noted that only the radial velocity component $u$ is perturbed, similarly as discussed in \cref{sec: SFBFS}.} We retain $R=30$ POD modes per trajectory for convergence analyses and the initial matrix $\boldsymbol{X}$ construction. To ensure reliable convergence, the matrix $\tilde{\boldsymbol{X}}$, from which the initial matrix $\boldsymbol{X}$ is constructed, is organised by ordering the POD modes. This systematic arrangement eliminates possible convergence issues arising from trajectory variations and results in smoother convergence curves, as shown in figure \ref{fig: 3dvortexcurves}$(a)$. Then the same methodology in \cref{LDP} is applied to conduct data-driven transient growth analyses at $\tau = 15, 30, 45, 60$.
	
	Figures \ref{fig: 3dvortexcurves}$(a)$, \ref{fig: 3dvortexcurves}$(b)$ and \ref{fig: 3dvortexcurves}$(c)$ effectively demonstrate the reliability of the data-driven methodology, even when applied to high-dimensional systems. Consistent with that in the backward-facing step case discussed in \cref{sec: SFBFS}, the optimal initial perturbations converge slightly more slowly than the optimal outcomes. Interestingly, the number of column vectors required in $\boldsymbol{X}$ is similar to that in \cref{sec: SFBFS}, despite the fact that the higher dimensionality typically demands more data. This suggests that the amount of data needed is not only strictly affected by the system's dimensionality but also influenced by the quality of the POD modes used to synthesise the optimal initial perturbations. This observation highlights the importance of well-chosen modes in ensuring an efficient and accurate representation of the system's dynamics.
	
	{We can also note that the error plots for optimal modes in figure \ref{fig: 3dvortexcurves}$(b)$ and \ref{fig: 3dvortexcurves}$(c)$ are non-monotonic. This is reasonable because when additional column vectors are included in $\boldsymbol{X}$, the corresponding POD subspace may change significantly, leading to a temporary misalignment between the reduced subspace and the true dominant transient dynamics. As shown in figure \ref{fig: vortexreview}, the azimuthal wavenumber of the computed optimal initial condition using the proposed method changes from 3 to 2 as the amount of data included in $\boldsymbol{X}$ increases. As an analogy, in model-based analyses, the computed optimal initial condition may only exhibit the correct wavenumber and structure after sufficient iterations. Thus, the observed oscillatory convergence pattern is an expected artefact when enriching the basis.}
	
	From the contours of the optimal initial condition at $\tau=15$ in figure \ref{fig: 3dvortextau1530}$(a)$, we observe that the approximation of the axial velocity is less accurate compared to those in radial and azimuthal directions. This discrepancy arises because the axial velocity is typically over two orders of magnitude smaller than the dominant swirl velocity. Additionally, the use of POD modes, which are prioritised based on their energy content, exacerbates this effect. From the contours of the optimal outcome, we see that the initially small radial and axial velocity components grow to become dominant, a phenomenon attributed to the so-called ``anti-lift-up” mechanism \citep{antkowiak2007vortex}. Figure \ref{fig: 3dvortexcurves}$(d)$ illustrates the energy evolution of optimal initial perturbations over time, which meets the optimal growth at the corresponding $\tau$. Therefore, the streamwise vorticity, related to the two dominant velocity components $u$ and $v$, is presented in figure \ref{fig: 3dvortextau1530}$(b)$ and the same results have also been shown in \citet{mao2012transient}. In contrast to the previous example, we observe that achieving mode convergence at later time instances is more challenging. This is because the out-of-core structure of the optimal initial conditions grows as the time instance increases, requiring a larger dataset to fully resolve these perturbations.

	\subsection {Correlation with DMD}\label{sec: Correlation with DMD}	
	{Although the method proposed in this work shares some similarities with DMD, the two have clear differences.}
	
	{DMD is primarily designed to capture the asymptotic behaviour of dynamical systems by approximating the system's long-time evolution. Considering the system in (\ref{DynamicEquation}), following most DMD-based methods \citep{schmid2010dynamic, tu2013dynamic, herrmann2021data}, transient growth analyses using DMD can be performed by first collecting snapshots from a set of trajectories into $\boldsymbol{X}_\text{DMD}=[ \boldsymbol{D}^1_{1:S-1}, \boldsymbol{D}^2_{1:S-1}, \ldots, \boldsymbol{D}^K_{1:S-1} ]$ and $\boldsymbol{Y}_\text{DMD}=[ \boldsymbol{D}^1_{2:S}, \boldsymbol{D}^2_{2:S}, \ldots, \boldsymbol{D}^K_{2:S} ]$, then a lower-dimensional surrogate $\hat{\boldsymbol{A}}$ with rank $r$ can be obtained via
		\[
		\begin{aligned}
			\boldsymbol{X}_\text{DMD} &= \boldsymbol{\mit{V}}_r \boldsymbol{\mit{\Lambda}}_r \boldsymbol{W}^{\mathrm{T}}_r, \\
			e^{\hat{\boldsymbol{A}} \Delta t} &= \boldsymbol{V}^{\mathrm{T}}_r \boldsymbol{Y}_\text{DMD} \boldsymbol{\mit{W}}_r \boldsymbol{\mit{\Lambda}}^{-1}_r,
		\end{aligned}
		\]
		where $\Delta t$ is the sampling interval. A possible solution to transient energy growth using DMD is 
		\[
		G(\tau)=\sigma_{\rm max} (\hat{\boldsymbol{F}}_\text{DMD} e^{\hat{\boldsymbol{A}} \tau} \hat{\boldsymbol{F}}^{-1}_\text{DMD}),
		\]
		where $\hat{\boldsymbol{F}}_\text{DMD}$ is determined by $\boldsymbol{V}^{\mathrm{T}}_r \boldsymbol{Q} \boldsymbol{V}_r = \hat{\boldsymbol{F}}_\text{DMD}^{\mathrm{T}} \hat{\boldsymbol{F}}_\text{DMD}$.}
	
	{In contrast, our method aims to accurately capture short-time transient dynamics by focusing on the finite-time evolution operator $\mathcal{M}(t)=e^{\boldsymbol{A} t}$, which is directly approximated by $\hat{\mathcal{M}}(t)$. However, in general, $\hat{\mathcal{M}}(t) \neq e^{\hat{\boldsymbol{A}} t}$, particularly in systems with strong non-normality, where spectral pollution leads to significant discrepancies between the evolution of the full and projected systems \citep{colbrook2024rigorous}.}
	
\begin{figure}
		\includegraphics[width=1\linewidth]{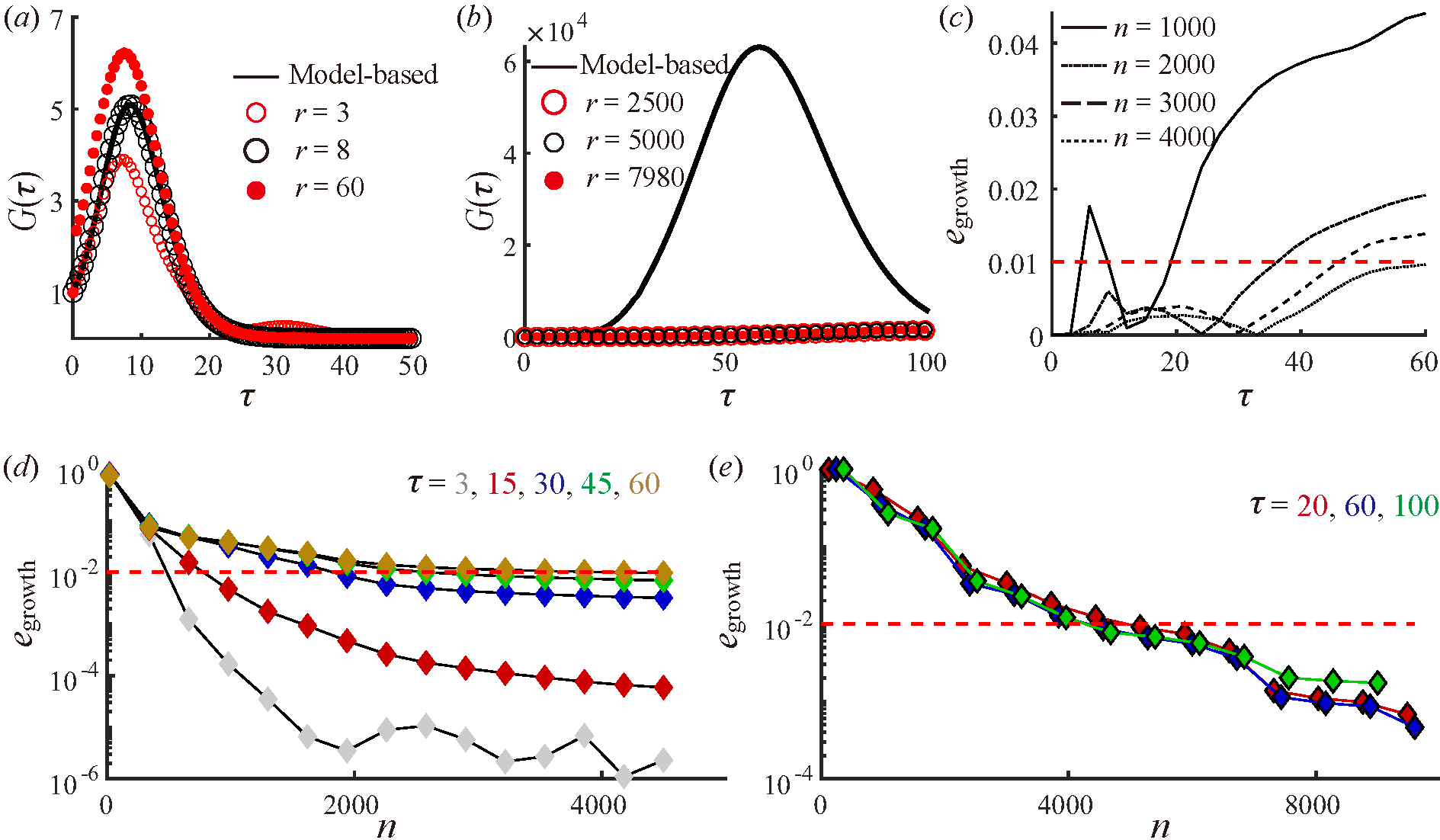}
		\captionsetup{width=1\textwidth}
		\caption{\justifying {$(a)$: Transient energy growth of the complex Ginzburg–Landau equation approximated using DMD with various rank $r$. $(b)$: Transient energy growth of the backward-facing step flow approximated using DMD with various rank $r$. $(c)$: Relative error of transient growth of the Batchelor vortex approximated by DMD, using dataset $\boldsymbol{X}$ and $\boldsymbol{Y}_{3}$ with different $n$. $(d)$: Relative error between model-based and data-driven transient growth at $\tau=3,15,30,45,60$ as a function of $n$ for the Batchelor vortex. $(e)$: Relative error between model-based and data-driven optimal energy growth at $\tau=20,60,100$ as a function of $n$ for the backward-facing step flow.}}
		\label{fig: 3DCurves_review}
	\end{figure}

	{To illustrate this, we conduct a numerical experiment of the complex Ginzburg–Landau equation using the same dataset as in \citet{herrmann2021data}. As shown in figure \ref{fig: 3DCurves_review}$(a)$, the DMD-based approximation is sensitive to the number of SVD modes retained. When the number of modes is insufficient, the maximum energy amplification is underestimated. Conversely, increasing the number of modes introduces spurious eigenmodes, which can result in overpredicted or non-physical energy growth. This highlights a key limitation of DMD-based techniques for accurately capturing transient amplification in non-normal systems and motivates the development of our direct evolution-operator-based framework.}
	
	{Although DMD-based technique demonstrates potential in the 1D case, its application to strongly non-normal systems remains challenging. For the backward-facing step flow, we collect snapshots of $\boldsymbol{X}_\text{DMD}$ and $\boldsymbol{Y}_\text{DMD}$ with $K=20$ and $S=400$, which are part of the data described in \cref{sec: SFBFS}. As shown in figure \ref{fig: 3DCurves_review}$(b)$, DMD-based technique fails to capture the transient energy growth, since the projected subspace spanned by $\boldsymbol{V}_r$ does not contain the optimal modes. This highlights the importance of the initial matrix construction discussed in \cref{sec: X construction}.}
	
	{Moreover, even if one chooses $\boldsymbol{X}_\text{DMD}=\boldsymbol{X}$ and $\boldsymbol{Y}_\text{DMD}=\boldsymbol{Y}_{\Delta t}$, the DMD-based approach remains ineffective to ensure accuracy at each time instance. As shown in figure \ref{fig: 3DCurves_review}$(c)$ and \ref{fig: 3DCurves_review}$(d)$, which presents an analysis of the Batchelor vortex, the energy growth prediction may converge at $\tau=3$, but this does not guarantee convergence of the maximum energy amplification at later times. For instance, at $\tau=60$, accurate prediction in this case requires retaining over 4000 column vectors in the initial matrix $\boldsymbol{X}$. This implies that to assess transient energy growth at $\tau=60$, we still need to seek a proper $n$ through integrating the vectors in $\boldsymbol{X}$ from $\tau=0$ to $\tau=60$, incurring a computational cost comparable to that of the full-order simulation described in our original method.} 
	
	{Last but not least, when we study the transient behaviour over the interval $\tau \in [0,{\tau}_{\max}]$, it is still not generally guaranteed that one can first compute a converged $\hat{\mathcal{M}}({\tau}_{\max})$ and then use $\hat{\mathcal{M}}(\Delta t)$ in the same dimension with $\hat{\mathcal{M}}({\tau}_{\max})$ to approximate the transient evolution at all intermediate times by $\hat{\mathcal{M}}(\tau)=[\hat{\mathcal{M}}(\Delta t)]^{\frac{\tau}{\Delta t}} $. While DMD-based approaches may appear feasible in certain cases, their success critically depends on knowing in advance the exact amount of data required for transient growth analysis over the interval $[0,{\tau}_{\max}]$. However, both the order of the column vectors in $\boldsymbol{X}$ and the spatial distribution of transient modes can significantly influence convergence. For example, in the case of flow over a backward-facing step, as shown in figure \ref{fig: 3DCurves_review}$(e)$, the amount of data required for computation at $\tau=20$ is larger than that at $\tau=60$ or $\tau=100$. This indicates that if the data needed most for transient growth analyses arises from a particular instant within $[0,{\tau}_{\max}]$, the more robust strategy is the one adopted in our method, i.e., directly collecting $\boldsymbol{Y}_{\tau}$ corresponding to each specific time $\tau$ of interest.}
	
	\section{Summary}\label{sec: Summary}	
	\setlength{\parskip}{0pt}
	In this work, we propose a data-driven methodology for transient growth analyses of linear dynamical systems and validate it through three case studies: the linearised complex Ginzburg–Landau equation, the flow over a backward-facing step, and the Batchelor vortex. The data-driven method utilises Hermite polynomials to gather transient information over short time intervals, followed by POD to extract dominant spatial modes from snapshots. By projecting the forward evolution operator onto a reduced space spanned by these modes, the data-driven method accurately captures the spatial distribution of the optimal initial perturbations, with their energy evolutions aligning with the model-based optimal growth. {The proposed algorithm also demonstrates robustness in avoiding the influence of spurious global modes, which are often encountered in DMD-based methods.}
	
	Furthermore, this study shows that optimal modes can also be interpreted as a combination of POD modes, which are chosen for the construction of the low-dimensional space in this method. In addition, the data-driven approach converges consistently as long as more data are included in the initial matrix $\boldsymbol{X}$ and outcome matrices $\boldsymbol{Y}_{\tau}$. The convergence behaviour also highlights the potential for more efficient strategies to further reduce the computational costs. 
	
	{A clear advantage of the proposed data-driven approach is that it eliminates the adjoint equations, relying solely on the forward operator/code to perform the analysis. This adjoint-free framework greatly simplifies the implementation and reduces the computational complexity. Contrasted with model-based methods, which require renewed forward and backward integrations for each time instance, the data-driven approach collects the snapshots at arbitrary time instance $t \in [0,\tau]$ during a single forward integration for $\boldsymbol{Y}_{\tau}$, facilitating the computation of transient growth over multiple time instances.} 
	
	While this data-driven method demonstrates superior performance for linearly stable flows, further investigations are needed for cases where nonlinear effects are significant or in compressible scenarios. Moreover, applying this method to data collected in experiments could be a valuable avenue for future research. {We also notice that the majority of wall time is dominated by I/O (reading/writing snapshots). Scalability for higher-dimensional problems can be improved by using parallelised I/O and GPU-accelerated linear algebra libraries.}

	\appendix
	
	\section{Hermite Polynomials}\label{HPolynomials}	
	
	\begin{figure}
		\includegraphics[width=1\linewidth]{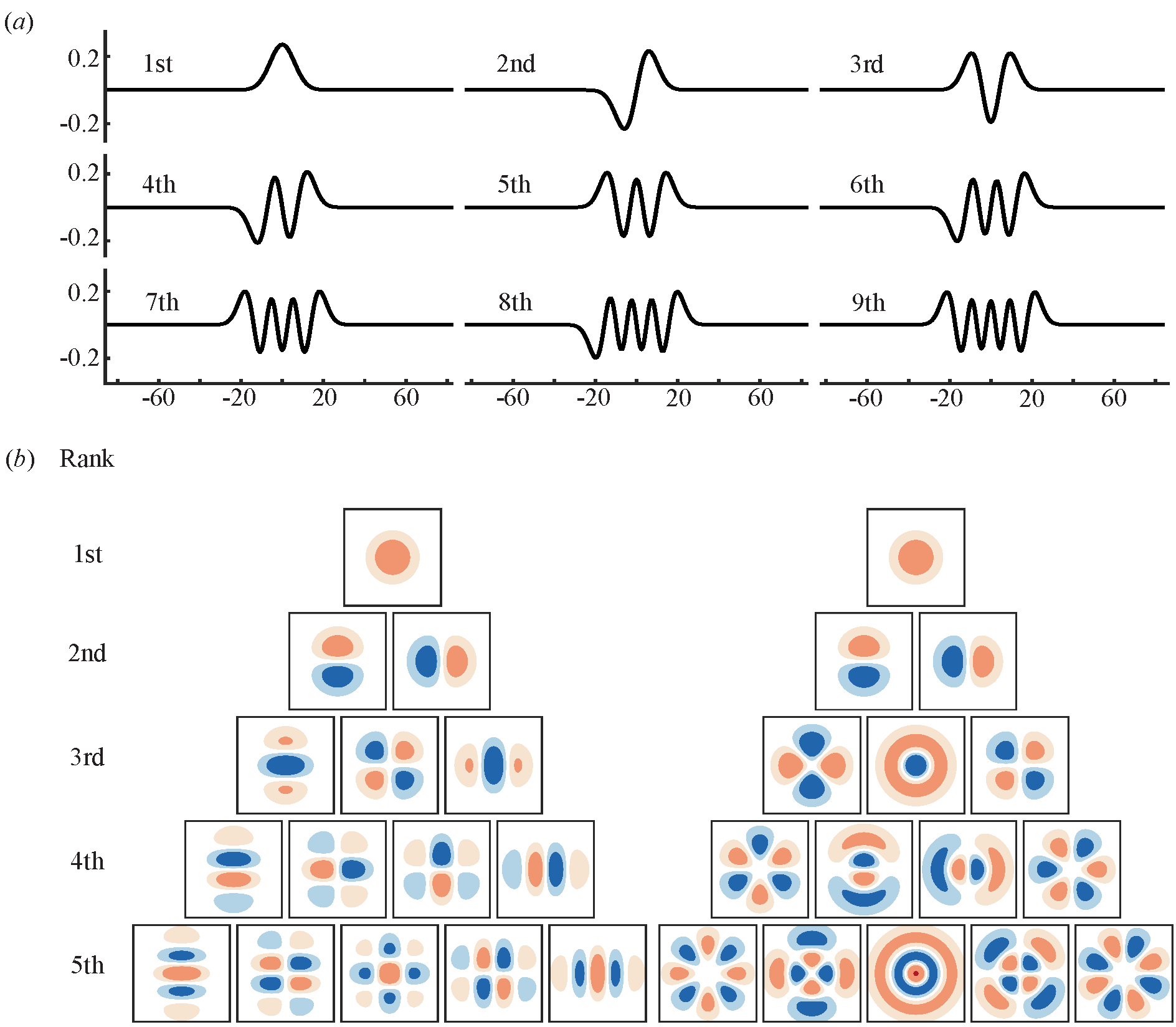}
		\captionsetup{width=1\textwidth}
		\caption{\justifying Hermite polynomials. $(a)$: The first $9$ Gauss-weighted Hermite functions. $(b)$: The first $5$ ranks of 2D Hermite polynomials in (left) Cartesian form and (right) polar form \citep{victor2006responses}.}
		\label{fig: 2dhermite}
	\end{figure}
	{The 1D Hermite polynomials satisfy the differential equation:
		\[
		\frac{d^2 h_{\rho}(x)}{dx^2} - \delta x \frac{d h_{\rho}(x)}{dx} + \delta n h_{\rho}(x) = 0,
		\]
		where $h_{\rho}(x)$ represents ${\rho}$th polynomial, which can also be defined by the relations below:
		\[
		\begin{aligned}
			h_1(x) &= 1, \\   
			h_2(x) &= \delta x, \\ 
			h_{m+2}(x) &= \delta x h_{m+1}(x) - m h_{m}(x) \quad \text { for } m = 1, 2, 3, \dots.
		\end{aligned}
		\]
		They satisfy the orthogonality with respect to the Gaussian weight $e^{-\frac{\delta}{2} x^{2}}$, namely
		\[
		\int_{-\infty}^{\infty} h_m(x) h_n(x)  e^{-\delta x^{2}} \mathrm{d} x=0 \quad \text { for } m \neq n,
		\]
		where $\delta > 0$ controls the decay rate of the Gaussian envelope. Those polynomials are complete in the Hilbert space, making them suitable for representing localised initial perturbations in unbounded domains. For enhanced adaptability within the computational domains, we employ Gauss-weighted Hermite functions with $\delta = 0.24$, given by 
		\[
		\Pi_{\rho} = h_{\rho}(x)\,e^{-\frac{\delta}{2} x^{2}}, 
		\]
		The first $9$ such functions are depicted in figure \ref{fig: 2dhermite}$(a)$.}
	
	The 2D Hermite polynomials can be constructed as products of 1D Hermite functions in the $x$ and $y$ directions \citep{victor2006responses}, satisfying the differential equation:
	\[
	\left( \frac{\partial}{\partial x} - \frac{\partial}{\partial y} \right)^{\alpha+\beta} H_{\alpha,\beta}(x,y) = 0.
	\]
	In Cartesian coordinates, the polynomials of rank $\rho=\alpha+\beta$, where $\rho \geq \alpha, \beta \geq 0$, are given by:
	\begin{equation}
		H_{\alpha, \beta}(x, y)=\frac{\Gamma}{C} h_\alpha \left(\frac{x}{C}\right) h_\beta \left(\frac{y}{C}\right) \exp \left(-\frac{x^2+y^2}{4 C^2}\right),
		\label{TDHc}
	\end{equation}
	where $h_\alpha(u)$ and $h_\beta(u)$ are 1D Hermite polynomials. $C$ represents the scale of the filter set and $\Gamma$ is a normalisation constant. When expressed in polar form, the 2D Hermite polynomials of the rank ${\rho} = \alpha + 2\beta$, where ${\rho} \geq \alpha \geq 0$ and $\alpha$, $\beta$ are positive integers, are structured into ``cosine'' and ``sine'' pairs:
	\begin{equation}
		\left\{\begin{aligned}
			H_{\alpha, \beta}^{\cos }(r, \theta)=\frac{\Gamma}{C} \cos (\alpha \theta)\left(\frac{r}{C}\right)^{\alpha} F_{\alpha, \beta}\left(\frac{r^2}{C^2}\right) \exp \left(-\frac{r^2}{4 C^2}\right); \\
			H_{\alpha, \beta}^{\sin }(r, \theta)=\frac{\Gamma}{C} \sin (\alpha \theta)\left(\frac{r}{C}\right)^{\alpha} F_{\alpha, \beta}\left(\frac{r^2}{C^2}\right) \exp \left(-\frac{r^2}{4 C^2}\right),
		\end{aligned}\right.
		\label{TDHp}
	\end{equation}
	with
	\[
	F_{p_c, p_s}(a)=\sum_{i=0}^{p_s}(-2)^{{p_s}-i} \frac{({p_c}+{p_s})!{p_s}!}{(p_c+i)!i!({p_s}-i)!} a^i.
	\]
	
	The first $5$ ranks of 2D Hermite polynomials in both polar and Cartesian coordinates are shown in figure \ref{fig: 2dhermite}$(b)$.
	
	\bibliographystyle{jfm}   
	\bibliography{jfm}   

\begin{thebibliography}{56}
\expandafter\ifx\csname natexlab\endcsname\relax\def\natexlab#1{#1}\fi
\def\au#1{#1} \def\ed#1{#1} \def\yr#1{#1}\def\at#1{#1}\def\jt#1{\textit{#1}}
  \def\bt#1{#1}\def\bvol#1{\textbf{#1}} \def\vol#1{#1} \def\pg#1{#1}
  \def\publ#1{#1}\def\arxiv#1{#1}\def\org#1{#1}\def\st#1{\textit{#1}}

\bibitem[Abdessemed {\em et~al.\/}(2009)Abdessemed, Sherwin \&
  Theofilis]{abdessemed2009linear}
{\sc \au{Abdessemed, Nadir}, \au{Sherwin, SJ} \& \au{Theofilis, Vassilios}}
  \yr{2009}  \at{Linear instability analysis of low-pressure turbine flows}.
  \jt{Journal of Fluid Mechanics}  \bvol{628},  \pg{57--83}.

\bibitem[Aboites \& Ram{\'\i}rez(2019)]{aboites2019simple}
{\sc \au{Aboites, Vicente} \& \au{Ram{\'\i}rez, Miguel}} \yr{2019}  \at{Simple
  approach to special polynomials: Laguerre, hermite, legendre, tchebycheff,
  and gegenbauer}.  \bt{In {\em Applied Mathematics\/}}.  \publ{IntechOpen}.

\bibitem[Andersson {\em et~al.\/}(1999)Andersson, Berggren \&
  Henningson]{andersson1999optimal}
{\sc \au{Andersson, Paul}, \au{Berggren, Martin} \& \au{Henningson, Dan~S}}
  \yr{1999}  \at{Optimal disturbances and bypass transition in boundary
  layers}.  \jt{Physics of Fluids}  \bvol{11}~(1),  \pg{134--150}.

\bibitem[Antkowiak \& Brancher(2007)]{antkowiak2007vortex}
{\sc \au{Antkowiak, Arnaud} \& \au{Brancher, Pierre}} \yr{2007}  \at{On vortex
  rings around vortices: an optimal mechanism}.  \jt{Journal of Fluid
  Mechanics}  \bvol{578},  \pg{295--304}.

\bibitem[Bagheri {\em et~al.\/}(2009)Bagheri, Henningson, Hoepffner \&
  Schmid]{bagheri2009input}
{\sc \au{Bagheri, Shervin}, \au{Henningson, Dan~S}, \au{Hoepffner, J} \&
  \au{Schmid, Peter~J}} \yr{2009}  \at{Input-output analysis and control design
  applied to a linear model of spatially developing flows} .

\bibitem[Bale \& Govindarajan(2010)]{bale2010transient}
{\sc \au{Bale, Rahul} \& \au{Govindarajan, Rama}} \yr{2010}  \at{Transient
  growth and why we should care about it}.  \jt{Resonance}  \bvol{15},
  \pg{441--457}.

\bibitem[Ballarin {\em et~al.\/}(2015)Ballarin, Manzoni, Quarteroni \&
  Rozza]{ballarin2015supremizer}
{\sc \au{Ballarin, Francesco}, \au{Manzoni, Andrea}, \au{Quarteroni, Alfio} \&
  \au{Rozza, Gianluigi}} \yr{2015}  \at{Supremizer stabilization of
  pod--galerkin approximation of parametrized steady incompressible
  navier--stokes equations}.  \jt{International Journal for Numerical Methods
  in Engineering}  \bvol{102}~(5),  \pg{1136--1161}.

\bibitem[Ballarin \& Rozza(2016)]{ballarin2016pod}
{\sc \au{Ballarin, Francesco} \& \au{Rozza, Gianluigi}} \yr{2016}
  \at{Pod--galerkin monolithic reduced order models for parametrized
  fluid-structure interaction problems}.  \jt{International Journal for
  Numerical Methods in Fluids}  \bvol{82}~(12),  \pg{1010--1034}.

\bibitem[Barkley {\em et~al.\/}(2008)Barkley, Blackburn \&
  Sherwin]{barkley2008direct}
{\sc \au{Barkley, Dwight}, \au{Blackburn, Hugh~Maurice} \& \au{Sherwin,
  Spencer~J}} \yr{2008}  \at{Direct optimal growth analysis for timesteppers}.
  \jt{International journal for numerical methods in fluids}  \bvol{57}~(9),
  \pg{1435--1458}.

\bibitem[Barkley {\em et~al.\/}(2002)Barkley, Gomes \&
  Henderson]{barkley2002three}
{\sc \au{Barkley, Dwight}, \au{Gomes, M Gabriela~M} \& \au{Henderson,
  Ronald~D}} \yr{2002}  \at{Three-dimensional instability in flow over a
  backward-facing step}.  \jt{Journal of fluid mechanics}  \bvol{473},
  \pg{167--190}.

\bibitem[Batchelor(1964)]{batchelor1964axial}
{\sc \au{Batchelor, GK173435}} \yr{1964}  \at{Axial flow in trailing line
  vortices}.  \jt{Journal of Fluid Mechanics}  \bvol{20}~(4),  \pg{645--658}.

\bibitem[Bellman(1966)]{bellman1966dynamic}
{\sc \au{Bellman, Richard}} \yr{1966}  \at{Dynamic programming}.  \jt{science}
  \bvol{153}~(3731),  \pg{34--37}.

\bibitem[Berkooz {\em et~al.\/}(1993)Berkooz, Holmes \&
  Lumley]{berkooz1993proper}
{\sc \au{Berkooz, Gal}, \au{Holmes, Philip} \& \au{Lumley, John~L}} \yr{1993}
  \at{The proper orthogonal decomposition in the analysis of turbulent flows}.
  \jt{Annual review of fluid mechanics}  \bvol{25}~(1),  \pg{539--575}.

\bibitem[Blackburn {\em et~al.\/}(2008)Blackburn, Barkley \&
  Sherwin]{blackburn2008convective}
{\sc \au{Blackburn, Hugh~Maurice}, \au{Barkley, Dwight} \& \au{Sherwin,
  Spencer~J}} \yr{2008}  \at{Convective instability and transient growth in
  flow over a backward-facing step}.  \jt{Journal of Fluid Mechanics}
  \bvol{603},  \pg{271--304}.

\bibitem[Colbrook(2023)]{colbrook2023multiverse}
{\sc \au{Colbrook, Matthew~J}} \yr{2023}  \at{The multiverse of dynamic mode
  decomposition algorithms}.  \jt{arXiv preprint arXiv:2312.00137} .

\bibitem[Colbrook \& Townsend(2024)]{colbrook2024rigorous}
{\sc \au{Colbrook, Matthew~J} \& \au{Townsend, Alex}} \yr{2024}  \at{Rigorous
  data-driven computation of spectral properties of koopman operators for
  dynamical systems}.  \jt{Communications on Pure and Applied Mathematics}
  \bvol{77}~(1),  \pg{221--283}.

\bibitem[Courant \& Hilbert(1962)]{courant2008methods}
{\sc \au{Courant, R.} \& \au{Hilbert, D.}} \yr{1962} {\em Methods of
  Mathematical Physics, Volume 2\/}.  \publ{Wiley}.

\bibitem[Cox \& Matthews(2002)]{cox2002exponential}
{\sc \au{Cox, Steven~M} \& \au{Matthews, Paul~C}} \yr{2002}  \at{Exponential
  time differencing for stiff systems}.  \jt{Journal of Computational Physics}
  \bvol{176}~(2),  \pg{430--455}.

\bibitem[Dawson {\em et~al.\/}(2016)Dawson, Hemati, Williams \&
  Rowley]{dawson2016characterizing}
{\sc \au{Dawson, Scott~TM}, \au{Hemati, Maziar~S}, \au{Williams, Matthew~O} \&
  \au{Rowley, Clarence~W}} \yr{2016}  \at{Characterizing and correcting for the
  effect of sensor noise in the dynamic mode decomposition}.  \jt{Experiments
  in Fluids}  \bvol{57},  \pg{1--19}.

\bibitem[De~Pando {\em et~al.\/}(2012)De~Pando, Sipp \&
  Schmid]{de2012efficient}
{\sc \au{De~Pando, Miguel~Fosas}, \au{Sipp, Denis} \& \au{Schmid, Peter~J}}
  \yr{2012}  \at{Efficient evaluation of the direct and adjoint linearized
  dynamics from compressible flow solvers}.  \jt{Journal of Computational
  Physics}  \bvol{231}~(23),  \pg{7739--7755}.

\bibitem[Deniz(2019)]{deniz2019quantum}
{\sc \au{Deniz, Co{\c{s}}kun}} \yr{2019}  \at{Quantum harmonic oscillator}.
  \bt{In {\em Oscillators-Recent Developments\/}}.  \publ{IntechOpen}.

\bibitem[van Dijk \& Martens(1996)]{van1996feature}
{\sc \au{van Dijk, Antoon~M} \& \au{Martens, J-B}} \yr{1996} Feature-based
  image compression with steered hermite transforms.  \bt{In {\em Proceedings
  of 3rd IEEE International Conference on Image Processing\/}}, ,
  \vol{vol.~1},  \pg{pp. 205--208}. IEEE.

\bibitem[Escalante-Ram{\'\i}rez \& L{\'o}pez-Caloca(2008)]{escalante2008multi}
{\sc \au{Escalante-Ram{\'\i}rez, Boris} \& \au{L{\'o}pez-Caloca, Alejandra~A}}
  \yr{2008} Multi-sensor image fusion with the steered hermite transform.
  \bt{In {\em Optical and Digital Image Processing\/}}, ,  \vol{vol. 7000},
  \pg{pp. 683--690}. SPIE.

\bibitem[Farrell(1988)]{farrell1988optimal}
{\sc \au{Farrell, Brian~F}} \yr{1988}  \at{Optimal excitation of perturbations
  in viscous shear flow}.  \jt{Physics of Fluids}  \bvol{31}~(8),  \pg{2093}.

\bibitem[Flinois {\em et~al.\/}(2015)Flinois, Morgans \&
  Schmid]{flinois2015projection}
{\sc \au{Flinois, Thibault~LB}, \au{Morgans, Aimee~S} \& \au{Schmid, Peter~J}}
  \yr{2015}  \at{Projection-free approximate balanced truncation of large
  unstable systems}.  \jt{Physical Review E}  \bvol{92}~(2),  \pg{023012}.

\bibitem[Girfoglio {\em et~al.\/}(2021)Girfoglio, Quaini \&
  Rozza]{girfoglio2021pod}
{\sc \au{Girfoglio, Michele}, \au{Quaini, Annalisa} \& \au{Rozza, Gianluigi}}
  \yr{2021}  \at{A pod-galerkin reduced order model for a les filtering
  approach}.  \jt{Journal of Computational Physics}  \bvol{436},  \pg{110260}.

\bibitem[Herrmann {\em et~al.\/}(2021)Herrmann, Baddoo, Semaan, Brunton \&
  McKeon]{herrmann2021data}
{\sc \au{Herrmann, Benjamin}, \au{Baddoo, Peter~J}, \au{Semaan, Richard},
  \au{Brunton, Steven~L} \& \au{McKeon, Beverley~J}} \yr{2021}  \at{Data-driven
  resolvent analysis}.  \jt{Journal of Fluid Mechanics}  \bvol{918},  \pg{A10}.

\bibitem[Hijazi {\em et~al.\/}(2020)Hijazi, Stabile, Mola \&
  Rozza]{hijazi2020data}
{\sc \au{Hijazi, Saddam}, \au{Stabile, Giovanni}, \au{Mola, Andrea} \&
  \au{Rozza, Gianluigi}} \yr{2020}  \at{Data-driven pod-galerkin reduced order
  model for turbulent flows}.  \jt{Journal of Computational Physics}
  \bvol{416},  \pg{109513}.

\bibitem[Karniadakis \& Sherwin(2005)]{karniadakis2005spectral}
{\sc \au{Karniadakis, George} \& \au{Sherwin, Spencer~J}} \yr{2005} {\em
  Spectral/hp element methods for computational fluid dynamics\/}.
  \publ{Oxford University Press, USA}.

\bibitem[Kutz {\em et~al.\/}(2016{\natexlab{{\em a\/}}})Kutz, Brunton, Brunton
  \& Proctor]{kutz2016dynamic}
{\sc \au{Kutz, J~Nathan}, \au{Brunton, Steven~L}, \au{Brunton, Bingni~W} \&
  \au{Proctor, Joshua~L}} \yr{2016{\natexlab{{\em a\/}}}} {\em Dynamic mode
  decomposition: data-driven modeling of complex systems\/}.  \publ{SIAM}.

\bibitem[Kutz {\em et~al.\/}(2016{\natexlab{{\em b\/}}})Kutz, Fu \&
  Brunton]{kutz2016multiresolution}
{\sc \au{Kutz, J~Nathan}, \au{Fu, Xing} \& \au{Brunton, Steven~L}}
  \yr{2016{\natexlab{{\em b\/}}}}  \at{Multiresolution dynamic mode
  decomposition}.  \jt{SIAM Journal on Applied Dynamical Systems}
  \bvol{15}~(2),  \pg{713--735}.

\bibitem[Landahl(1980)]{landahl1980note}
{\sc \au{Landahl, MT}} \yr{1980}  \at{A note on an algebraic instability of
  inviscid parallel shear flows}.  \jt{Journal of Fluid Mechanics}
  \bvol{98}~(2),  \pg{243--251}.

\bibitem[Lumey(2012)]{lumey2012stochastic}
{\sc \au{Lumey, John~L}} \yr{2012} {\em Stochastic tools in turbulence\/}.
  \publ{Elsevier}.

\bibitem[Mao(2015)]{mao2015effects}
{\sc \au{Mao, X}} \yr{2015}  \at{Effects of base flow modifications on noise
  amplifications: flow past a backward-facing step}.  \jt{Journal of fluid
  mechanics}  \bvol{771},  \pg{229--263}.

\bibitem[Mao {\em et~al.\/}(2013)Mao, Blackburn \& Sherwin]{mao2013calculation}
{\sc \au{Mao, Xuerui}, \au{Blackburn, Hugh~M} \& \au{Sherwin, Spencer~J}}
  \yr{2013}  \at{Calculation of global optimal initial and boundary
  perturbations for the linearised incompressible navier--stokes equations}.
  \jt{Journal of Computational Physics}  \bvol{235},  \pg{258--273}.

\bibitem[Mao \& Sherwin(2012)]{mao2012transient}
{\sc \au{Mao, X} \& \au{Sherwin, SJ}} \yr{2012}  \at{Transient growth
  associated with continuous spectra of the batchelor vortex}.  \jt{Journal of
  fluid mechanics}  \bvol{697},  \pg{35--59}.

\bibitem[Marisa {\em et~al.\/}(2015)Marisa, Niederhauser, Haeberlin, Wildhaber,
  Vogel, Jacomet \& Goette]{marisa2015bufferless}
{\sc \au{Marisa, Thanks}, \au{Niederhauser, Thomas}, \au{Haeberlin, Andreas},
  \au{Wildhaber, Reto~A}, \au{Vogel, Rolf}, \au{Jacomet, Marcel} \& \au{Goette,
  Josef}} \yr{2015}  \at{Bufferless compression of asynchronously sampled ecg
  signals in cubic hermitian vector space}.  \jt{IEEE Transactions on
  Biomedical Engineering}  \bvol{62}~(12),  \pg{2878--2887}.

\bibitem[Marquet \& Sipp(2012)]{marquet2012convective}
{\sc \au{Marquet, O} \& \au{Sipp, D}} \yr{2012}  \at{Convective instabilities
  in a backward-facing step flow: global forced perturbations}.  \jt{Progress
  in Flight Physics}  \bvol{3},  \pg{451--460}.

\bibitem[Murata {\em et~al.\/}(2020)Murata, Fukami \&
  Fukagata]{murata2020nonlinear}
{\sc \au{Murata, Takaaki}, \au{Fukami, Kai} \& \au{Fukagata, Koji}} \yr{2020}
  \at{Nonlinear mode decomposition with convolutional neural networks for fluid
  dynamics}.  \jt{Journal of Fluid Mechanics}  \bvol{882},  \pg{A13}.

\bibitem[Nekkanti \& Schmidt(2021)]{nekkanti2021frequency}
{\sc \au{Nekkanti, Akhil} \& \au{Schmidt, Oliver~T}} \yr{2021}
  \at{Frequency--time analysis, low-rank reconstruction and denoising of
  turbulent flows using spod}.  \jt{Journal of Fluid Mechanics}  \bvol{926},
  \pg{A26}.

\bibitem[Obrist \& Schmid(2003)]{obrist2003linear}
{\sc \au{Obrist, Dominik} \& \au{Schmid, Peter~J}} \yr{2003}  \at{On the linear
  stability of swept attachment-line boundary layer flow. part 2. non-modal
  effects and receptivity}.  \jt{Journal of Fluid Mechanics}  \bvol{493},
  \pg{31--58}.

\bibitem[Raposo {\em et~al.\/}(2019)Raposo, Mughal \&
  Ashworth]{raposo2019adjoint}
{\sc \au{Raposo, Henrique}, \au{Mughal, Shahid} \& \au{Ashworth, Richard}}
  \yr{2019}  \at{An adjoint compressible linearised navier--stokes approach to
  model generation of tollmien--schlichting waves by sound}.  \jt{Journal of
  Fluid Mechanics}  \bvol{877},  \pg{105--129}.

\bibitem[Reddy \& Henningson(1993)]{reddy1993energy}
{\sc \au{Reddy, Satish~C} \& \au{Henningson, Dan~S}} \yr{1993}  \at{Energy
  growth in viscous channel flows}.  \jt{Journal of Fluid Mechanics}
  \bvol{252},  \pg{209--238}.

\bibitem[Sayadi \& Schmid(2016)]{sayadi2016parallel}
{\sc \au{Sayadi, Taraneh} \& \au{Schmid, Peter~J}} \yr{2016}  \at{Parallel
  data-driven decomposition algorithm for large-scale datasets: with
  application to transitional boundary layers}.  \jt{Theoretical and
  Computational Fluid Dynamics}  \bvol{30},  \pg{415--428}.

\bibitem[Schmid(2007)]{schmid2007nonmodal}
{\sc \au{Schmid, Peter~J}} \yr{2007}  \at{Nonmodal stability theory}.
  \jt{Annu. Rev. Fluid Mech.}  \bvol{39}~(1),  \pg{129--162}.

\bibitem[Schmid(2010)]{schmid2010dynamic}
{\sc \au{Schmid, Peter~J}} \yr{2010}  \at{Dynamic mode decomposition of
  numerical and experimental data}.  \jt{Journal of fluid mechanics}
  \bvol{656},  \pg{5--28}.

\bibitem[Schmid(2022)]{schmid2022dynamic}
{\sc \au{Schmid, Peter~J}} \yr{2022}  \at{Dynamic mode decomposition and its
  variants}.  \jt{Annual Review of Fluid Mechanics}  \bvol{54}~(1),
  \pg{225--254}.

\bibitem[Schmid \& Henningson(2001)]{schmidpj2001stabilityandtransitionin}
{\sc \au{Schmid, Peter~J} \& \au{Henningson, Dan~S}} \yr{2001} Stability and
  transition in shear flows.

\bibitem[Sirovich(1987)]{sirovich1987turbulence}
{\sc \au{Sirovich, Lawrence}} \yr{1987}  \at{Turbulence and the dynamics of
  coherent structures. i. coherent structures}.  \jt{Quarterly of applied
  mathematics}  \bvol{45}~(3),  \pg{561--571}.

\bibitem[Trefethen {\em et~al.\/}(1993)Trefethen, Trefethen, Reddy \&
  Driscoll]{trefethen1993hydrodynamic}
{\sc \au{Trefethen, Lloyd~N}, \au{Trefethen, Anne~E}, \au{Reddy, Satish~C} \&
  \au{Driscoll, Tobin~A}} \yr{1993}  \at{Hydrodynamic stability without
  eigenvalues}.  \jt{Science}  \bvol{261}~(5121),  \pg{578--584}.

\bibitem[Tu(2013)]{tu2013dynamic}
{\sc \au{Tu, Jonathan~H}} \yr{2013}  \at{Dynamic mode decomposition: Theory and
  applications}. PhD thesis, Princeton University.

\bibitem[Tumin \& Reshotko(2001)]{tumin2001spatial}
{\sc \au{Tumin, Anatoli} \& \au{Reshotko, Eli}} \yr{2001}  \at{Spatial theory
  of optimal disturbances in boundary layers}.  \jt{Physics of Fluids}
  \bvol{13}~(7),  \pg{2097--2104}.

\bibitem[Victor {\em et~al.\/}(2006)Victor, Mechler, Repucci, Purpura \&
  Sharpee]{victor2006responses}
{\sc \au{Victor, Jonathan~D}, \au{Mechler, Ferenc}, \au{Repucci, Michael~A},
  \au{Purpura, Keith~P} \& \au{Sharpee, Tatyana}} \yr{2006}  \at{Responses of
  v1 neurons to two-dimensional hermite functions}.  \jt{Journal of
  neurophysiology}  \bvol{95}~(1),  \pg{379--400}.

\bibitem[Wei \& Rowley(2009)]{wei2009low}
{\sc \au{Wei, Mingjun} \& \au{Rowley, Clarence~W}} \yr{2009}
  \at{Low-dimensional models of a temporally evolving free shear layer}.
  \jt{Journal of Fluid Mechanics}  \bvol{618},  \pg{113--134}.

\bibitem[Williams {\em et~al.\/}(2015)Williams, Kevrekidis \&
  Rowley]{williams2015data}
{\sc \au{Williams, Matthew~O}, \au{Kevrekidis, Ioannis~G} \& \au{Rowley,
  Clarence~W}} \yr{2015}  \at{A data--driven approximation of the koopman
  operator: Extending dynamic mode decomposition}.  \jt{Journal of Nonlinear
  Science}  \bvol{25},  \pg{1307--1346}.

\bibitem[Wynn {\em et~al.\/}(2013)Wynn, Pearson, Ganapathisubramani \&
  Goulart]{wynn2013optimal}
{\sc \au{Wynn, Andrew}, \au{Pearson, DS}, \au{Ganapathisubramani, Bharathram}
  \& \au{Goulart, Paul~J}} \yr{2013}  \at{Optimal mode decomposition for
  unsteady flows}.  \jt{Journal of Fluid Mechanics}  \bvol{733},
  \pg{473--503}.

\end{thebibliography}
\end{document}